\documentclass[english,10pt,aps,prd,eqsecnum,nofootinbib]{revtex4}
\usepackage{amssymb,amsmath,amsthm,graphicx,amscd}
\usepackage[dvipsnames]{xcolor}
\usepackage{esint,enumerate,comment,ulem}
\usepackage{babel}

\begin{document}

\title{Nonequilibrium Nonlinear Open Quantum Systems I. \\  Functional Perturbative Analysis of a Weakly Anharmonic Oscillator}

\author{Jen-Tsung Hsiang$^{1,2}$}%
\thanks{cosmology@gmail.com}%
\affiliation{$^{1}$ Center for High Energy and High Field Physics, National Central University, Chungli 32001, Taiwan}
\affiliation{$^{2}$ Center for Particle Physics and Field Theory, Department of Physics, Fudan University, Shanghai 200438, China}

\author{Bei-Lok Hu$^{3}$}%
\thanks{blhu@umd.edu}
\affiliation{$^{3}$Maryland Center for Fundamental Physics and Joint Quantum Institute, \\ University of Maryland, College Park, Maryland 20742, USA}

\begin{abstract}
We introduce a functional perturbative method for treating weakly nonlinear systems coupled with a quantum field bath.  We demonstrate using this method to obtain the covariance matrix elements and the correlation functions of a quantum anharmonic oscillator interacting with a heat bath. We  identify a fluctuation-dissipation relation based on the nonequilibrium dynamics of this nonlinear open quantum system. To establish its connection with dynamical equilibration, we further examine the energy flows between the anharmonic oscillator and the bath field.  The vanishing of the net flow {is an indication of} the existence of an equilibrium state for such an open-system configuration. The  results presented here are useful for studying the nonequilibrium physical processes of nonlinear quantum systems such as heat transfer or electron transport. 
\end{abstract}
\maketitle

\setcounter{tocdepth}{1}

\baselineskip=18pt

\allowdisplaybreaks

\section{Introduction}

In this series of papers we study the nonequilibrium dynamics of nonlinear open quantum systems, using the quantum Brownian oscillator~\cite{Sch61,FeyVer63,CalLeg83,GSI88,UnrZur89,HPZ92} as a generic model.  For this first paper (I) we  use the functional perturbative method originated in~\cite{HPZ93} to study the dynamics of an anharmonic oscillator linearly coupled to a thermal field bath. In a follow-up paper (II)~\cite{NLFDR} {we provide a nonperturbative derivation of the fluctuation-dissipation relation (FDR) of this nonlinear system at late times under specified assumptions/conditions (e.g. in the nonchaotic regime). We then use perturbative results for the weakly anharmonic oscillator open quantum system to  analyze these conditions. More details will be provided after this brief background description.

\paragraph{Nonlinear} While linear analysis has occupied the attention of physicists and mathematicians for the longest time in theoretical science we know very well that large scale structures of Nature are more aptly and accurately depicted by nonlinear systems. Likewise, while the equilibrium condition is often assumed and most studied in statistical mechanics and thermodynamics, it cannot capture the real time dynamics of a many-body system.  Nonequilibrium statistical physics is needed.  Studies of the nonequilibrium dynamics of nonlinear systems in the last half century have spawned new fields, such as driven diffusive and dissipative systems. This is further enriched when noise and stochastic processes are incorporated --  the interplay of nonlinearity and stochasticity reveals many interesting phenomena such as those pertaining to forms and growth, and created new fields, such as stochastic thermodynamics, etc.  

\paragraph{Quantum} Studies of the nonequilibrium properties of quantum systems are rapidly gaining importance as quantum devices are becoming smaller, the functional numbers (e.g., of photon) are becoming fewer.  New resources in quantum coherence and entanglement also enable  innovations with wide ranging applications  in quantum science and technology, notably quantum information processing. 

\paragraph{Nonequilibrium} Real time measurements of quantum systems are now becoming increasingly feasible. Experimental developments have stimulated further advances in the theoretical studies of the nonequilibrium dynamics of quantum open systems. Quantum Brownian motion (QBM) is probably the best known paradigm in the description of quantum dissipative and stochastic processes.  For linear systems our understanding of QBM has progressed from Markovian processes {\it a la} the Lindblad-GKS master equations~\cite{Lindblad,GKS} in the 60s-70s, the Caldeira-Leggett master equation~\cite{CalLeg83} in the 80s to nonMarkovian processes {\it a la} the Hu-Paz-Zhang master equation~\cite{HPZ92} in the 90s. These foundational work have invigorated  new disciplines like open quantum systems~\cite{qos,CalHu2008} and enriched new fields such as  quantum thermodynamics~\cite{qtdbooks,QTD}. They also serve as an effective platform to construct theories of nonlinear open quantum systems, the goal of this series of papers.

There are many lines of inquiry into the fundamental issues of  quantum and statistical physics of nonlinear quantum open systems.  We mention two examples here:  1) Quantum decoherence~\cite{qdec}, 2) Quantum chaos~\cite{qchaos}. Using the decoherent history formalism~\cite{dechis}, following what was done earlier~\cite{qcldomain} for quasi-classical domains  with linear Brownian motion models, Brun~\cite{Brun} derived the quasiclassical equations of motion for nonlinear Brownian systems.  The case of linear interactions was treated exactly, and nonlinear interactions are also compared, using classical and quantum perturbation theory.  On the issue of  quantum chaos and quantum-classical transition, the systematic work of Habib's group is noteworthy~\cite{Habib}.  For the most recent developments of quantum chaos involving OTOC, see e.g., the work of Ueda's group~\cite{Ueda} (where earlier references of AdS-CFT relevance can be found).

\paragraph{Advantages of the functional method}

In this paper we use the functional perturbative method developed in~\cite{HPZ93} for the study of a weakly anharmonic oscillator linearly coupled to a thermal field bath. With a driven quantum oscillator linearly coupled to an environment as the unperturbed system, we can find the physical observables in the nonlinear system by taking the suitable functional derivatives of the counterparts in the unperturbed system with respect to the driving source. In~\cite{HPZ93} this method was applied to the cases  where the coupling of the system with the environment is (weakly) nonlinear. Here we treat the case where the system itself is (weakly) nonlinear, but keep the oscillator-bath coupling linear. {As a necessary preparation procedure}, we first calculate the quantum statistical averages of the operators of the covariance matrix elements and then use the functional method to introduce real-time functions  for the driven, harmonic (linear) oscillator.  These results serve as the base or background from which we begin calculating the  higher-order (nonlinear) corrections via the functional perturbative approach. In particular, we use the four-point function of the anharmonic oscillator and its first-order correction as an illustration of the functional method. For an arbitrary Gaussian initial state of the oscillator the functional method enables one to write the multi-point functions and their higher-order corrections in terms of the zeroth-order two-point functions in the form of a Wick-like expansion in a fairly straightforward way. By contrast, in the operator approach this procedure can be dauntingly complicated since there are four types of real-time two-point functions\footnote{Since they can be related by one equation only three of them are independent.} in the `in-in' (Schwinger-Keldysh or closed-time-path) formalism~\cite{Sch61,ctp}. Finding the appropriate combinations of these two-point functions for the higher-order results can take considerable effort in the canonical operator approach. {(Notwithstanding, its advantages will be exploited in a different way described below.)}  These combinations of the zeroth-order two-point functions reveal the hidden structures of the higher-order results, and allow us to find the late time behavior and identify a fluctuation-dissipation relation (FDR) of the oscillator for the first-order corrections in the anharmonicity.  

\paragraph{Dynamical equilibration and FDR}

For the linear (harmonic) oscillator it has been shown~\cite{QTD,HHL18,AFM0} that dynamical equilibration is a necessary and sufficient condition for the existence of a FDR. When the equilibrium state is reached, the energy exchange between the oscillator and the thermal bath will be balanced. The opposite is also true. In the current configuration, to ask if dynamical equilibration is consummated, we have calculated the energy flows between the anharmonic oscillator and the bath quantum field order by order. We show that up to first-order in anharmonicity, each channel of energy flow indeed becomes time-independent and their sum vanishes, indicating a balance of the energy exchange between the anharmonic oscillator and the bath field. {This is clearly seen from the real time evolution of this nonlinear open quantum system, such as shown in Figs.~\ref{Fi:PowEx} and \ref{Fi:PowBal}. Nonequilibrium dynamics shows the net average energy flow in the anharmonic oscillator--quantum field system decreases in time and eventually comes to a halt.} We further examine the response of the first-order correction of the energy flow at late times for the oscillator in contact with the bath which is initially at zero or a high temperature. The correction does not depend on the temperature of the bath when it is sufficiently high, as seen in Fig.~\ref{Fi:PowHT}. {It implies that equilibration as indicated by the net energy flow goes on for a wide range of bath temperatures from low to high.}  We also derive an FDR at late times between the first-order corrections of the noise and the dissipation kernels. Interestingly, it has the same functional form as the FDR for the zeroth-order dynamics. On one hand, it implies a connection between equilibration and FDR at late times even for the anharmonic oscillator. {On the other hand, these results inspire  a \textit{nonperturbative} derivation of the fluctuation-dissipation relation for the anharmonic oscillator in the open quantum systems framework, which is shown in a follow-up paper }(II)~\cite{NLFDR}. There, from the nonperturbative expressions of the powers delivered from the environment  to the anharmonic oscillator, and dissipated out of it back to the environment, we show, under some additional assumptions, that the existence of a stable equilibrium state {warrants} a FDR, thus highlighting the connection between stable dynamical equilibration and {the appearance of an FDR}. A future paper (III)~\cite{NENL3} in this series studies two and more weakly nonlinearly coupled quantum oscillators with the end oscillators coupled linearly to two baths of different temperatures. We  {will} examine the late time nonequilibrium steady-state behavior of this system  by the functional perturbative method {introduced in this paper to analytically address the validity of Fourier law in energy transport in nonlinear quantum systems.}  Results from this study are useful for the investigation of heat transfer or charge transport in these nonlinear quantum open systems.

\paragraph{Functional perturbative approach}
 
In principle, we can obtain the expectation values of the linear system once we have the density operator\footnote{For a Gaussian system, its density matrix operator can be quite straightforwardly written down because in the coordinate representation it is Gaussian. The exponent is a second-order polynomial of the canonical coordinates and their coefficients can be expressed in terms of the expectation values of the covariance matrix elements. However, in this construction, there is a chicken-or-egg dilemma: you do not know the density matrix unless you know the expectation values of the covariance matrix elements, but you do not know the expectation values of the covariance matrix elements unless you have the density matrix. Thus we still need to use other approaches to break open the loop.} of such a system and then compute the trace of the product of the observable operator and the density operator. However, this procedure, which involves performing integrations, can be quite tedious in evaluating the expectation values of the various elements of the covariance matrix. A similar but  more efficient approach is to calculate only the in-in generating functional of the linear system, driven by external sources. The expectation values of the covariance matrix elements or the two-point functions of the canonical variables can then be found by taking the functional derivatives. We will focus on this approach but still offer the explicit calculations based on the density matrix in the Appendices for comparison. Similarly, the correction to the two-point functions, or the expectation values of quantum observables due to the nonlinear potential can also be found order by order by taking the appropriate functional derivatives of the generating functional of the linear system linked to external sources, a commonly employed technique in  quantum field theory. On the other hand, the canonical operator approach, {notwithstanding the aforementioned inconvenience in the higher-order calculations, has a distinct advantage: Due to their resemblance to the classical equations of motion the coupled Heisenberg equations are more intuitive. They provide a clear physical picture of the dynamics of this coupled system, and allow us to identify the physical observables in the context more easily. We will thus use the operator description for these purposes.}


\section{Driven Linear Oscillators: Nonequilibrium Evolution}\label{S:drnkw}
To illustrate how the functional techniques are applied to nonlinear open quantum systems it is convenient to first deal with the case of a linear oscillator, driven by an external source $j$, bilinearly coupled to a quantum scalar field. Taking the functional derivatives with respect to this external source one can derive quantities associated with the linear oscillator, and in perturbative orders, the nonlinear oscillator. Thus we begin with a description of the nonequilibrium evolution of a linear oscillator driven by an external source $j$. 

  The  action for a harmonic oscillator of mass $m$ and bare natural frequency $\omega_{0}$ coupled to a  massless Klein-Gordon field $\phi$ in $1+3$ dimensional Minkowski spacetime takes on the form
\begin{align}\label{E:fkgbrkjts}
	 S[\chi,\phi]&=\int_{0}^{t}\!ds\;\Bigl\{\frac{m}{2}\Bigl[\dot{\chi}^{2}(s)-\omega_{0}^{2}\chi^{2}(s)\Bigr]+j(s)\chi(s)\Bigr\}\\
	 &\qquad\qquad\qquad+\int^{t}_{0}\!d^{4}x\;e\chi(s)\delta^{3}(\mathbf{x}-\mathbf{z}(s))\phi(x)+\int^{t}_{0}\!d^{4}x\;\frac{1}{2}\,\bigl[\partial_{\mu}\phi(x)\bigr]\bigl[\partial^{\mu}\phi(x)\bigr]\Bigr\}\,,\notag
\end{align}
where in the dipole approximation 
$\chi$ is decoupled from its external trajectory $\mathbf{z}(s)$.  Because the whole system is Gaussian we can solve the problem exactly with no need to assume small coupling strength  $e$   between the oscillator and the field. We allow the {oscillator} to move along a prescribed trajectory $\mathbf{z}(s)$, and thus this action includes that which describes the Unruh effect~\cite{Urn76} whence the linear oscillator is referred to as  {the internal degree of freedom of} an Unruh-DeWitt detector. Note the difference between the oscillator's idf $\chi$ and external degree of freedom (edf) $\mathbf{z}(s)$. For our purpose here, $\chi$ is also driven by an external source $j$, and $\mathbf{z}$ is chosen to be a constant.  

We now derive the the in-in generating functional for this driven linear open quantum system. Perturbative treatment of driven nonlinear open quantum systems will be treated in the next section.

\subsection{Driven harmonic oscillator}
The reduced density operator of a free harmonic oscillator linearly coupled with a scalar field bath has been derived before~\cite{GSI88,HHAoP,HHgrdsta}. The corresponding reduced density operator for the driven harmonic oscillator can be formally obtained in the same manner,
\begin{align}\label{E:ngbrhtdfs}
	 \rho_{\chi}(\chi_{b},\chi'_{b};t)&=\int_{-\infty}^{\infty}\!d\chi_{a}d\chi'_{a}\;\rho_{\chi}(\chi_{a},\chi'_{a},0)\!\int_{\chi_{a}}^{\chi_{b}}\!\mathcal{D}\chi_{+}\!\int_{\chi'_{a}}^{\chi'_{b}}\!\mathcal{D}\chi_{-}\;\exp\Bigl(i\,S_{\chi}[\chi_{+}]-i\,S_{\chi}[\chi_{-}]+i\,S_{I}[\chi_{+}]-i\,S_{I}[\chi_{-}]\Bigr)\notag\\
	 &\qquad\qquad\quad\times\exp\biggl\{\frac{i}{2}\,e^{2}\int_{0}^{t}\!ds\,ds'\biggl(\Bigl[\chi_{+}(s)-\chi_{-}(s)\Bigr]G_{R}^{(\phi)}(s,s')\Bigl[\chi_{+}(s')+\chi_{-}(s')\Bigr]\biggr.\biggr.\notag\\
	 &\qquad\qquad\qquad\qquad\qquad\qquad\quad\;+\biggl.\biggl.i\,\Bigl[\chi_{+}(s)-\chi_{-}(s)\Bigr]G_{H,\,\beta}^{(\phi)}(s,s')\Bigl[\chi_{+}(s')-\chi_{-}(s')\Bigr]\biggr)\biggr\}\,,
\end{align}
where
\begin{align*}
	S_{\chi}[\chi]&=\int_{0}^{t}\!ds\;\frac{m}{2}\Bigl[\dot{\chi}^{2}(s)-\omega_{0}^{2}\chi^{2}(s)\Bigr]\,,&S_{I}[\chi]&=\int_{0}^{t}\!ds\;j(s)\chi(s)\,,
\end{align*}
and $\rho_{\chi}(\chi_{a},\chi'_{a},0)$ is the density matrix of the oscillator at the initial time $t_{a} = 0$. The two kernel functions $G_{R}^{(\phi)}(s,s')$ and $G_{H,\beta}^{(\phi)}(s,s')$ are respectively the retarded Green's function and the Hadamard function of the  {free} scalar field evaluated at spacetime points $(s,\mathbf{z})$ and $(s',\mathbf{z})$. They are defined by
\begin{align}
	G_{R}^{(\phi)}(s,s')&=i\,\theta(s-s')\,\operatorname{Tr}_{\phi}\Bigl(\rho_{\beta}^{(\phi)}({0})\,\bigl[\phi(s,\mathbf{z}),\,\phi(s',\mathbf{z})\bigr]\Bigr)\,,\\
	G_{H}^{(\phi)}(s,s')&=\frac{1}{2}\,\operatorname{Tr}_{\phi}\Bigl(\rho_{\beta}^{(\phi)}( {0})\,\bigl\{\phi(s,\mathbf{z}),\,\phi(s',\mathbf{z})\bigr\}\Bigr)\,,
\end{align}
if the initial state of the scalar field is a thermal state $\rho_{\beta}^{(\phi)}(t_{a})$ at the temperature $\beta^{-1}$. The function $\theta(\tau)$ is the unit-step function.

\subsubsection{Coarse-grained effective action }
Now introduce the center-of-mass coordinate $r$ and the relative coordinate $q$
\begin{align*}
	r&=\frac{\chi_{+}+\chi_{-}}{2}\,,&q&=\chi_{+}-\chi_{-}\,,&j_{r}&=\frac{j_{+}+j_{-}}{2}\,,&j_{q}&=j_{+}-j_{-}\,,
\end{align*}
the reduced density operator becomes
\begin{align}\label{E:kdjwiqa}
	 \rho_{\chi}(r_{b},q_{b};t)&=\int_{-\infty}^{\infty}\!dr_{a}dq_{a}\;\rho_{\chi}(r_{a},q_{a},0)\!\int_{r_{a}}^{r_{b}}\!\mathcal{D}r\!\int_{q_{a}}^{q_{b}}\!\mathcal{D}q\;\exp\Bigl(i\,S_{CG}[r,q]\Bigr)\,,
\end{align}
with the coarse-grained effective action $S_{CG}$~\cite{cgea} given by
\begin{align}\label{E:lkqoakj}
	 S_{CG}[r,q]&=\int_{0}^{t}\!ds\;\biggl\{m\Bigl[\dot{q}(s)\dot{r}(s)-\omega_{0}^{2}\,q(s)r(s)\Bigr]+r(s)j_{q}(s)+q(s)j_{r}(s)\biggr.\notag\\
	 &\qquad\qquad\qquad+\biggl.e^{2}\int_{0}^{t}\!ds'\;\Bigl[q(s)\,G_{R}^{(\phi)}(s,s')\,r(s')+\frac{i}{2}\,q(s)\,G_{H,\,\beta}^{(\phi)}(s,s')\,q(s')\Bigr]\biggr\}\,.
\end{align}
This effective action describes all the influences from the environmental field in the form of convolution integrals in terms of the kernel functions.

Since the integrand of the path integrals in \eqref{E:kdjwiqa} is Gaussian, the path integration will give a result proportional the exponential of the classical value $\overline{S}_{CG}$ of the coarse-grained effective action
\begin{equation}\label{E:gkjrtsf}
	\rho_{\chi}(q_{b},r_{b};t)=\mathcal{N}\int\!dq_{a}\,dr_{a}\;\exp\Bigl[i\,\overline{S}_{CG}\Bigr]\rho_{\chi}(q_{a},r_{a};0)\,,
\end{equation}
where $\overline{S}_{CG}$, evaluated in Appendix~\ref{S:btiete}, is given by
\begin{equation}
	\overline{S}_{CG}[r,q]=m\,q(s)\dot{r}_{R}(s)\,\Big|_{s=0}^{s=t}+\frac{i}{2}\,e^{2}\int_{0}^{t}\!ds\int_{0}^{t}\!ds'\;q(s)\,G_{H,\,\beta}^{(\phi)}(s,s')\,q(s')+\int_{0}^{t}\!ds\;r_{}(s)j_{q}(s)\,.
\end{equation}
The functional forms of $r(s)$ and $q(s)$ can be found in \eqref{E:uierda} and \eqref{E:vierda},  and $r_{R}(s)$ denotes the real part of $r(s)$.
 
If we take the trace of the final reduced density operator, it gives an expression which is the in-in version of the generating functional  {at time $t$} (different from  the more familiar in-out version):
\begin{align}
	\mathcal{Z}[j_{q},j_{r}{;t)}&=\operatorname{Tr}\Bigl\{\rho_{\chi}(q_{b},r_{b};t)\Bigr\}=\mathcal{N}\int\!dq_{b}\,dr_{b}\int\!dq_{a}\,dr_{a}\;\delta(q_{b})\,\exp\Bigl[i\,\overline{S}_{CG}\Bigr]\rho_{\chi}(q_{a},r_{a};0)\,.\label{E:kowuqha}
\end{align}
The normalization $\mathcal{N}$ determined by $\mathcal{Z}[0,0]=1$ takes on the value
\begin{equation}
	\mathcal{N}=\frac{m}{2\pi\,D_{2}(t)}\,.
\end{equation}
The function $D_{2}(t)$, {as well as $D_{1}(t)$} given in Appendix~\ref{S:btiete}, is related to the retarded Green's function of the linear damped oscillator.

\subsubsection{`In-In' generating functional $\mathcal{Z}$ for driven linear systems}\label{S:bgerthsb}
It proves useful to have an explicit form of the generating functional $\mathcal{Z}[j_{q},j_{r}]$ when the initial state of the oscillator has a Gaussian form. Starting from \eqref{E:kowuqha}, we have
\begin{align*}
	 \mathcal{Z}[j_{q},j_{r};t)&=\frac{m}{2\pi\,D_{2}(t)}\int\!dq_{b}\,dr_{b}\;\delta(q_{b})\int\!dq_{a}\,dr_{a}\;\rho_{\chi}(q_{a},r_{a};0)\,\exp\left\{i\,m\,q(s)\dot{r}(s)\,\Big|_{s=0}^{s=t}\right.\notag\\
	 &\qquad\qquad\qquad\qquad\qquad\qquad\qquad-\left.\frac{e^{2}}{2}\int_{0}^{t}\!ds\int_{0}^{t}\!ds'\;q(s)\,G_{H,\,\beta}^{(\phi)}(s,s')\,q(s')+i\int_{0}^{t}\!ds\;r_{}(s)j_{q}(s)\right\}\,.
\end{align*}
If, for example, the initial state of the oscillator is a wavepacket of width $\sigma$
\begin{equation}\label{E:eubkjfg}
	 \rho_{\chi}(q_{a},r_{a},0)=\left(\frac{1}{\pi\sigma^{2}}\right)^{1/2}\exp\biggl\{-\frac{1}{\sigma^{2}}\left[r_{a}^{2}+\frac{1}{4}\,q_{a}^{2}\right]\biggr\}\,,
\end{equation}
then carrying out the Gaussian integrals for the generating functional $\mathcal{Z}[j_{q},j_{r}]$ gives
\begin{align}\label{E:ngmshtse}  
	 \mathcal{Z}[j_{q},j_{r};t) &=\exp\biggl\{-\frac{1}{4}\int_{0}^{t}\!ds\!\int_{0}^{t}\!ds'\;j_{q}(s)\Bigl[\sigma^{2}D_{1}(s)D_{1}(s')+\frac{1}{m^{2}\sigma^{2}}\,D_{2}(s)D_{2}(s')\Bigr]j_{q}(s')\biggr.\\
	 &\qquad\qquad+\biggl.\frac{i}{m}\int_{0}^{t}\!ds\!\int_{0}^{s}\!ds'\;j_{q}(s)\,D_{2}(s-s')\,j_{r}(s')-\frac{e^{2}}{2}\int_{0}^{t}\!ds\!\int_{0}^{t}\!ds'\;\mathfrak{J}_{q}(s)\,G_{H,\,\beta}^{(\phi)}(s-s')\,\mathfrak{J}_{q}(s')\biggr\}\,.\notag
\end{align}
where
\begin{equation}
	\mathfrak{J}_{q}(s)=\frac{1}{m}\int_{0}^{t}\!ds'\;D_{2}(s'-s)\,j_{q}(s')
\end{equation}
Details of these derivations are given in Appendix~\ref{S:kgbrkt}. Note that when $j_{q}=0=j_{r}$, we indeed have $\mathcal{Z}[0,0]$=1. 

{This generating functional serves as the starting point of the functional approach to  deriving the nonequilibrium dynamics of a nonlinear oscillator  coupled to a scalar quantum field bath.  Before proceeding, as an example, we want to show how to use the functional method to derive the two-point functions of the linear oscillator. They are necessary building blocks in the functional perturbative approach to tackle the anharmonic oscillator open system.}

\subsection{Quantum statistical averages of operators in linear systems}\label{S:brkfbdrt}

With the explicit expression for the reduced density matrix elements $\rho_{\chi}(q_{b},r_{b};t)$ of the driven linear oscillator at any time $t$, we may straightforwardly find the quantum (expectation value) statistical average  $\langle \hat{O}(t) \rangle_{j} $ of the operator associated with a physical variable ${O}$ at that time.  
For example, the quantum statistical average of  $\hat{\chi}_b=\hat{\chi}(t)$ is formally given by
\begin{equation}\label{E:dgbkre}
 \langle\,\hat{\chi}_{b}\,\rangle_{j} =\frac{1}{\mathcal{Z}[j]}\int\!dq_{b}\,dr_{b}\int\!dq_{a}\,dr_{a}\;\delta(q_{b})
 \left[r_{b}+\frac{1}{2}\,q_{b}\right]\,\rho_{\chi}(q_{b},r_{b};t)\,.
\end{equation}
We keep a subscript $j$ to remind ourselves of the dependence on the external source. This allows us to find the quantum statistical averages of physical variables associated with the nonlinear oscillator by performing {further} functional differentiations, as will be outlined in Sec.~\ref{S:ernkw}.

 However, this can be quite tedious and repetitive if the reduced density operator is very complicated, and can pose a real challenge for the perturbative calculations. Fortunately these tasks can be accomplished in a more economical way by taking the functional derivatives of the generating functional $\mathcal{Z}$ \eqref{E:ngmshtse} with respect to the external sources.

{The covariance matrix elements, and various real-time two-point functions of the linear oscillator form the core structure from which one can deduce the nonequilibrium quantum dynamics of a Gaussian system~\cite{QTD,HHAoP,HHgrdsta,HHL18,HH15} and its quantum informational properties such as the entanglement measures~\cite{Adesso}. Same is true for the weakly nonlinear system based on a perturbative treatment using the Gaussian system as its (zeroth order) base. Thus, as a start, it may be helpful to illustrate how to  use the functional method to find these quantities with a few elementary examples. In these cases, the functional method may look like an overkill, as they can be and have been obtained by other approaches. However, as we delve deeper into nonlinear systems seeking higher-order corrections,  the advantage of the functional method will become more apparent. Roughly speaking, in higher-order calculations of the nonequilibrium systems via perturbation theory, we often find it convenient to express the results in terms of the real-time two-point functions of the zeroth (linear or harmonic) order. As we mentioned earlier in the description of the in-in formalism, there are four types of Green's function for the system of interest, instead of the sole Feynman propagator in the in-out formalism. Thus performing the Wick-like expansion in the higher-order calculations in terms of suitable combinations of these four Green's functions for a general initial Gaussian state can be a formidable task for, say, the operator approach, but it is quite straightforward via the functional method. Examining and explaining this procedure by a simpler example can help to identify the underlying structures in the higher-order results, and thus facilitates the realization of the FDR with the perturbative treatment discussed in the following section, and by the nonperturbative argument in~\cite{NLFDR}.}

\subsubsection{Covariance matrix element $\langle\,\hat{\chi}^{2}(\tau)\,\rangle$}

We demonstrate this route by showing how to compute one element, $\langle\,\hat{\chi}^{2}(\tau)\,\rangle$, the second moment of the displacement of the free harmonic oscillator coupled to the field. {Appendix~\ref{S:eibde} includes a few more examples.}

It is useful to first introduce a couple of shorthand notations
\begin{align}
	\frac{\delta\mathcal{Z}[j;t)}{\delta j_{q}(\tau)}&=i\,\Xi[j;\tau)\,\mathcal{Z}[j;t)\,,\qquad\qquad\qquad\qquad\qquad\qquad\frac{\delta\mathcal{Z}[j;t)}{\delta j_{r}(\tau)}=i\,\mathfrak{J}_{q}(\tau)\,\mathcal{Z}[j;t)\,,\label{E:ngjset}
\intertext{and}
	\Xi[j;\tau)&=\frac{i}{2}\int_{0}^{t}\!ds'\;\Bigl[\sigma^{2}D_{1}(\tau)D_{1}(s')+\frac{1}{m^{2}\sigma^{2}}\,D_{2}(\tau)D_{2}(s')\Bigr]j_{q}(s')\biggr.\\
	&\qquad\qquad\qquad+\biggl.\frac{1}{m}\int_{0}^{t}\!ds'\;D_{2}(\tau-s')\,j_{r}(s')+i\,\frac{e^{2}}{m}\int_{0}^{t}\!ds\!\int_{0}^{t}\!ds'\;D_{2}(\tau-s)\,G_{H,\,\beta}^{(\phi)}(s-s')\,\mathfrak{J}_{q}(s')\,,\notag
\end{align} 
for $0<\tau<t$. We use the collective index $j$ to represent the pair of external currents $(j_{+},j_{-})$ or $(j_{q},j_{r})$, depending on the context.

We thus find
\begin{align}\label{E:dbsrjht}
	\frac{1}{\mathcal{Z}[j;t)}\frac{\delta^{2}\mathcal{Z}[j;t)}{i^{2}\delta j^{2}_{+}(\tau)}&=-\frac{1}{\mathcal{Z}[j;t)}\left\{\frac{\delta^{2}\mathcal{Z}[j;t)}{\delta j^{2}_{q}(\tau)}+\frac{\delta^{2}\mathcal{Z}[j;t)}{\delta j_{q}(\tau)\,\delta j_{r}(\tau)}+\frac{1}{4}\frac{\delta^{2}\mathcal{Z}[j;t)}{\delta j^{2}_{r}(\tau)}\right\}\,.
\end{align}
On the righthand side, only
\begin{align}
	\frac{\delta\,\Xi[j;\tau)}{\delta j_{q}(\tau)}&=\frac{i}{2}\biggl[\sigma^{2}D^{2}_{1}(\tau)+\frac{1}{m^{2}\sigma^{2}}\,D^{2}_{2}(\tau)\biggr]+i\,\frac{e^{2}}{m^{2}}\int_{0}^{\tau}\!ds'\!\int_{0}^{\tau}\!ds''\;D_{2}(\tau-s')G_{H,\,\beta}^{(\phi)}(s'-s'')D_{2}(\tau-s'')\,,
\end{align}
does not vanish in the limit $j\to 0$. {Eq.~\eqref{E:dbsrjht} then gives}
\begin{align}
	\langle\,\hat{\chi}_{+}^{2}(\tau)\,\rangle_{j}&=-\frac{1}{\mathcal{Z}[j;t)}\frac{\delta^{2}\mathcal{Z}[j;t)}{\delta j^{2}_{+}(\tau)}=\frac{1}{\mathcal{Z}[j;t)}\left\{\frac{\delta^{2}\mathcal{Z}[j;t)}{\delta j^{2}_{q}(\tau)}+\frac{\delta^{2}\mathcal{Z}[j;t)}{\delta j_{q}(\tau)\,\delta j_{r}(\tau)}+\frac{1}{4}\frac{\delta^{2}\mathcal{Z}[j;t)}{\delta j^{2}_{r}(\tau)}\right\}\,.
\end{align}
{It can be shown that taking the functional derivatives with respect to $j_{-}$ alone or a combination of $j_{+}$ and $j_{-}$ will give the same result} 
\begin{align}
	\langle\,\hat{\chi}_{b}^{2}\,\rangle&=\frac{1}{2}\left[\sigma^{2}D_{1}^{2}(t)+\frac{1}{m^{2}\sigma^{2}}\,D_{2}^{2}(t)\right]+\frac{e^{2}}{m^{2}}\int_{0}^{t}\!ds\!\int_{0}^{t}\!ds'\;D_{2}(t-s)\,G_{H,\,\beta}^{(\phi)}(s-s')\,D_{2}(t-s')\,,
\end{align}
in the limit $j\to0$ because the time-ordering is irrelevant. We also note that it is identical to the result
obtained by the reduce density matrix {in Appendix~\ref{S:eibde}}. {Next, we would like to show how to use the functional method to derive the two-point functions of the linear oscillator. They are necessary building blocks in the functional perturbative approach to tackle the anharmonic oscillator open system.}

\subsection{Real-time two-point functions of the linear system}\label{S:rbkher}

According to the `in-in' formalism~\cite{Sch61} in terms of the closed-time path integrals~\cite{ctp}, we may introduce path-ordered two-point functions by
\begin{align}\label{E:fkjtbfdgf}
	\langle\,\mathcal{P}\hat{\chi}(\tau)\hat{\chi}(\tau')\,\rangle=\begin{cases}
		\langle\,\mathcal{T}\hat{\chi}(\tau)\hat{\chi}(\tau')\,\rangle\,,&\tau\in C_{+}\;\&\;\tau'\in C_{+}\,,\\
		\langle\,\hat{\chi}(\tau')\hat{\chi}(\tau)\,\rangle\,,&\tau\in C_{+}\;\&\;\tau'\in C_{-}\,,\\
		\langle\,\hat{\chi}(\tau)\hat{\chi}(\tau')\,\rangle\,,&\tau\in C_{-}\;\&\;\tau'\in C_{+}\,,\\
		\langle\,\mathcal{T}^{*}\hat{\chi}(\tau)\hat{\chi}(\tau')\,\rangle\,,&\tau\in C_{-}\;\&\;\tau'\in C_{-}\,,
	\end{cases}
\end{align}
 where $C_{+/-}$ represents the forward/backward time branch and $\mathcal{T}$, $\mathcal{T}^{*}$ denote time-ordering and anti-time-ordering.  {We will adapt this protocol to the quantum open systems in order to identify the two-point functions of the reduced system. This will be justified once we re-write the in-in generating functional~\eqref{E:ngmshtse} by these two-point functions, to be discussed later.}

\subsubsection{Feynman propagator $\langle\,\mathcal{T}\,\hat{\chi}(\tau)\hat{\chi}(\tau')\,\rangle$}
Let us first find the Feynman propagator by evaluating the second derivatives of the generating functional with respect to $j_{+}$ at two different times, that is, with $0<\tau,\,\tau'<t$,
\begin{align}
	\langle\,\mathcal{T}\,\hat{\chi}(\tau)\hat{\chi}(\tau')\,\rangle&=-\frac{1}{\mathcal{Z}[j;t)}\frac{\delta^{\,2}\mathcal{Z}[j;t)}{\delta j_{+}(\tau)\,\delta j_{+}(\tau')}\;\bigg|_{\substack{j=0\\q=0}}\label{E:tbdfse}\\
	&=-\frac{1}{\mathcal{Z}[j;t)}\left\{\frac{\delta^{2}\mathcal{Z}[j;t)}{\delta j_{q}(\tau)\,\delta j_{q}(\tau')}+\frac{1}{2}\frac{\delta^{2}\mathcal{Z}[j;t)}{\delta j_{q}(\tau')\delta j_{r}(\tau)}+\frac{1}{2}\frac{\delta^{2}\mathcal{Z}[j;t)}{\delta j_{q}(\tau)\,\delta j_{r}(\tau')}+\frac{1}{4}\frac{\delta^{2}\mathcal{Z}[j;t)}{\delta j_{r}(\tau)\,\delta j_{r}(\tau')}\right\}\;\bigg|_{\substack{j=0\\q=0}}\notag\\
	&=-i\,\frac{\delta\,\Xi[j;\tau')}{\delta j_{q}(\tau)}-\frac{i}{2m}\,D_{2}(\tau'-\tau)-\frac{i}{2m}\,D_{2}(\tau-\tau')+\cdots\;\bigg|_{\substack{j=0\\q=0}}\,,\label{E:eruhwimbcd}
\end{align}
where $\cdots$ represents terms that will vanish in the $j\to0$ limit. Compare \eqref{E:eruhwimbcd} with the decomposition of the Feynman propagator of the free Klein Gordon field $\phi$, 
\begin{equation}
	\langle\,\mathcal{T}\hat{\phi}(x)\hat{\phi}(x')\,\rangle=G_{H}^{(\phi)}(x,x')-\frac{i}{2}\Bigl[G_{R}^{(\phi)}(x,x')+G_{A}^{(\phi)}(x,x')\Bigr]\,,
\end{equation} 
where $G_{A}^{(\phi)}(x,x')=G_{R}^{(\phi)}(x',x)$ is the corresponding advanced Green's function.  {Since this decomposition of the expectation value of the time-ordered product of operator holds quite generally}, we can make these identifications
\begin{align}
	G_{H}^{(\chi)}(\tau,\tau')&=-i\,\frac{\delta\,\Xi[j;\tau')}{\delta j_{q}(\tau)}\label{E:bgfhke1}\\
	&=\frac{1}{2}\biggl[\sigma^{2}D_{1}(\tau)D_{1}(\tau')+\frac{1}{m^{2}\sigma^{2}}\,D_{2}(\tau)D_{2}(\tau')\biggr]+\frac{e^{2}}{m^{2}}\int_{0}^{\tau}\!ds'\!\int_{0}^{\tau'}\!ds''\;D_{2}(\tau-s')G_{H,\,\beta}^{(\phi)}(s'-s'')D_{2}(\tau'-s'')\,,\notag
\intertext{and}
	G_{R}^{(\chi)}(\tau,\tau')&=\frac{1}{m}\,D_{2}(\tau-\tau')\,,\label{E:bgfhke2}
\end{align}
{for the operator $\hat{\chi}$ of the reduced system}. It is consistent with the fact that $D_{2}(\tau)$ is a form of retarded function according to its definition in terms of the Laplace transformation.

 {It is interesting to note that with the identifications of \eqref{E:bgfhke1} and \eqref{E:bgfhke2}, the in-in generating functional of the  {driven} harmonic oscillator \eqref{E:ngmshtse} can be written as
\begin{align}
	\mathcal{Z}[j]&=\exp\biggl\{i\int_{0}^{t}\!ds\int_{0}^{s}\!ds'\;j_{q}(s)\,G_{R,0}^{(\chi)}(s-s')\,j_{r}(s')-\frac{1}{2}\int_{0}^{t}\!ds\int_{0}^{t}\!ds'\;j_{q}(s)\,G_{H,0}^{(\chi)}(s,s')\,j_{q}(s')\biggr\}\label{E:ggsebssd}\\
	&=\exp\biggl\{\frac{i}{2}\oint_{C}\!ds\oint_{C}\!ds'\;j(s)\,G_{P,0}^{(\chi)}(s,s')\,j(s')\biggr\}\,,\label{E:dbeirdsf}
\end{align}
where a condensed notion is introduced in the second line, namely, $C=C_{+}+C_{-}$ and $j(s)=\pm j_{\pm}(s)$ when $s\in C_{\pm}$. The closed contour runs from $s=0$ to $s=t$ along the branch $C_{+}$ and then back to $s=0$ along the branch $C_{-}$. The path-ordered Green's function $G_{P,0}^{(\chi)}(s,s')$ is defined analogous to the time-ordered Green's function, according to the locations of $s$, $s'$ along the closed contour $C$. Note that the recognition of \eqref{E:dbeirdsf} makes connections with the Keldysh formalism and it can be viewed as an extension of the Keldysh formalism to the open systems. Most of all, it can greatly simplify the higher-order calculations, such as for the four-point functions. However, since this generating functional is obtained for the reduced system in the framework of nonequilibrium open systems, the path-ordered two-point functions in \eqref{E:dbeirdsf} in general is not stationary even though we have chosen a stationary initial state for the reduced system.}

In this section we show that when the linear oscillator is coupled to a scalar quantum field, by taking suitable functional derivatives of the `in-in' generating functional of the oscillator driven by the external sources, we can find the time evolution of the covariance matrix elements, and the real-time two-point functions associated with the free linear oscillator. It can be less painstaking and more efficient than the corresponding calculations based on the reduced density matrix of the free linear oscillator when it couples with the field. Similar results can also obtained by the canonical approach as exemplified in~\cite{LinHu09}. 
These results will be further extended to the nonlinear oscillator.


\section{Nonequilibrium Anharmonic Oscillator}\label{S:ernkw}

In the previous section we have calculated the expectation values associated with a free harmonic oscillator in contact with a thermal bath, modeled by a quantum scalar field initially at $t=0$ in a thermal state. Here we wish to treat the nonequilibrium evolution of an anharmonic oscillator based on the perturbative functional approach. We also would like to calculate the corresponding expectation values  for the anharmonic oscillator.

Suppose the action takes the form
\begin{align} \label{DrivenNL}
	 S[\chi,\phi;j]&=\int_{0}^{t}\!ds\;\Bigl\{\frac{m}{2}\Bigl[\dot{\chi}^{2}(s)-\omega_{0}^{2}\chi^{2}(s)\Bigr]-V[\chi(s)]+j(s)\chi(s)\Bigr\}\notag\\
	 &\qquad\qquad\qquad+\int^{t}_{0}\!d^{4}x\;e\chi(s)\delta^{3}(\mathbf{x}-\mathbf{z}(s))\phi(x)+\int^{t}_{0}\!d^{4}x\;\frac{1}{2}\,\bigl[\partial_{\mu}\phi(x)\bigr]\bigl[\partial^{\mu}\phi(x)\bigr]\Bigr\}\,,\\
	&=S_{0}[\chi,\phi;j]-\int_{0}^{t}\!ds\;V(\chi)\,.
\end{align}
Here the nonlinear potential is represented by $V(\chi)$ and contains a small parameter. The final reduced density matrix $\rho^{(V)}_{\chi}(t)$ is then given by
\begin{align}
	 \rho_{\chi}^{(V)}(q_{b},r_{b},t)&=\int\!dq_{a}\,dr_{a}\!\int_{q_{a}}^{q_{b}}\!\mathcal{D}q\int_{r_{a}}^{r_{b}}\!\mathcal{D}r\;\exp\biggl\{-i\int_{0}^{t}\!ds\;\Bigl[V(r+\frac{1}{2}\,q)-V(r-\frac{1}{2}\,q)\Bigr]\biggr\}\notag\\
	&\qquad\qquad\qquad\qquad\times\exp\biggl\{i\,S_{CG}[r,q;j_{q},j_{r}]\biggr\}\rho_{\chi}(q_{a},r_{a},0)\\
	&=\exp\biggl\{-i\int_{0}^{t}\!ds\;\Bigl[V(\frac{\delta}{i\,\delta j_{+}})-V(-\frac{\delta}{i\,\delta j_{-}})\Bigr]\biggr\}\,\rho_{\chi}(q_{b},r_{b},t)\;\bigg|_{j_{q}=0=j_{r}}\,,
\end{align}
where the density matrix element $\rho_{\chi}(q_{b},r_{b},t)$ of the free oscillator has been given by \eqref{E:gkjrtsf} and
\begin{align}
	\frac{\delta}{\delta j_{+}}&=+\frac{\delta}{\delta j_{q}}+\frac{1}{2}\frac{\delta}{\delta j_{r}}\,,&\frac{\delta}{\delta j_{-}}&=-\frac{\delta}{\delta j_{q}}+\frac{1}{2}\frac{\delta}{\delta j_{r}}\,.
\end{align}
Thus the expectation value of an operator $\hat{\mathcal{O}}$  consisting only the system variables is given by
\begin{align}
	 \langle\,\hat{\mathcal{O}}\,\rangle&=\frac{1}{\mathcal{Z}_{V}}\,\operatorname{Tr}\Bigl\{\mathcal{O}\,\rho^{(V)}_{\chi}(t_{b})\Bigr\}=\frac{1}{\mathcal{Z}_{V}}\exp\biggl\{-i\int_{0}^{t}\!ds\;\Bigl[V(\frac{\delta}{i\,\delta j_{+}})-V(-\frac{\delta}{i\,\delta j_{-}})\Bigr]\biggr\}\langle\,\hat{\mathcal{O}}\,\rangle_{0}\,\mathcal{Z}_{}\;\bigg|_{j_{q}=0=j_{r}}\,,
\end{align}
with $\mathcal{Z}=\operatorname{Tr}\{\rho_{\chi}(t)\}$, $\mathcal{Z}_{V}=\operatorname{Tr}\{\rho_{\chi}^{(V)}(t)\}$ being the `in-in' generating {functionals of the driven oscillators in the absence or the presence of the anharmonic potential, respectively,} at the final time $t$ to ensure proper normalization.  The subscript $0$ on an expectation value $\langle\,\mathcal{O}\,\rangle_{0}$ refers  to the results without the anharmonic potential.

For example, consider the anharmonic potential
\begin{align}\label{E:ukbwzaf}
	V(\chi)&=\frac{\lambda}{3!}\,\chi^{3}\,,\qquad\qquad&\lambda&>0\,,
\end{align}
where the self-coupling strength $\lambda$ is assumed to be small. This configuration inherently can induce dynamical instability; however, {in the context of perturbation theory, we consider only small amplitude motion about $\chi=0$ such that the mechanical energy of the oscillator remains much lower than the local maximum of the potential energy. Moreover,} interaction with the scalar field environment renders the system the configuration of a damped nonlinear oscillator driven by the quantum fluctuations of the field. 
When the stochastic force associated with the quantum noise is sufficiently weak, the dissipation tends to make the oscillator stably confined within a small range of the local minimum of the potential, so that the higher-order contributions may remain small even in the long time limit.

The functional derivatives associated with the nonlinear potential is explicitly given by
\begin{align}
	V(\frac{\delta}{i\,\delta j_{+}})-V(-\frac{\delta}{i\,\delta j_{-}})&=\frac{\lambda}{3!}\biggl\{3\left(\frac{\delta}{i\,\delta j_{q}}\right)^{2}\left(\frac{\delta}{i\,\delta j_{r}}\right)+\frac{1}{4}\left(\frac{\delta}{i\,\delta j_{r}}\right)^{3}\biggr\}\,.
\end{align}
Thus the leading order nonlinear contribution to the expectation value from the nonlinear potential is
\begin{equation}\label{E:gkhers}
	 \langle\,\hat{\mathcal{O}}\,\rangle^{(1)}=-i\,\frac{\lambda}{\mathcal{Z}_{V}}\int_{0}^{t}\!ds\;\biggl\{\frac{1}{2!}\left(\frac{\delta}{i\,\delta j_{q}(s)}\right)^{2}\left(\frac{\delta}{i\,\delta j_{r}(s)}\right)+\frac{1}{4!}\left(\frac{\delta}{i\,\delta j_{r}(s)}\right)^{3}\biggr\}\,\langle\,\hat{\mathcal{O}}\,\rangle_{0}\,\mathcal{Z}_{}\;\bigg|_{j_{q}=0=j_{r}}\,.
\end{equation}
Next we apply this to find the corrections to the in-in generating functional.

\subsection{Generating functional $\mathcal{Z}_{V}$ for driven nonlinear systems}

The full generating functional $\mathcal{Z}_{V}$ in the presence of nonlinear action can be expanded by
\begin{align}
	\mathcal{Z}_{V}=\exp\biggl\{-i\int_{0}^{t}\!ds\;\Bigl[V(\frac{\delta}{i\,\delta j_{+}})-V(-\frac{\delta}{i\,\delta j_{-}})\Bigr]\biggr\}\,\mathcal{Z}&=\mathcal{Z}+\mathcal{Z}_{1}+\cdots\,,
\end{align}
where $\mathcal{Z}_{1}$ is the leading order correction of $\mathcal{Z}$ due to the nonlinear potential $V$. Thus $\mathcal{Z}_{1}$ is given by
\begin{align}
	 \mathcal{Z}_{1}&=-i\,\frac{\lambda}{\mathcal{Z}_{V}}\int_{0}^{t}\!ds\;\biggl\{\frac{1}{2!}\left(\frac{\delta}{i\,\delta j_{q}(s)}\right)^{2}\left(\frac{\delta}{i\,\delta j_{r}(s)}\right)+\frac{1}{4!}\left(\frac{\delta}{i\,\delta j_{r}(s)}\right)^{3}\biggr\}\,\mathcal{Z}[j;t)\;\bigg|_{j_{q}=0=j_{r}}\,.
\end{align}
First, we show from \eqref{E:ngjset} that
\begin{align}
	\frac{\delta^{3}\mathcal{Z}[j;t)}{\delta j^{3}_{r}(\tau)}&=-i\,\mathfrak{J}^{3}_{q}(\tau)\,\mathcal{Z}[j;t)\,,&\frac{\delta^{3}\mathcal{Z}[j;t)}{\delta j^{2}_{q}(\tau)\,\delta j^{}_{r}(\tau)}&=-\mathfrak{J}_{q}(\tau)\,\frac{\delta}{\delta j^{}_{q}(\tau)}\,\biggl\{\Xi[j;\tau)\mathcal{Z}[j;t)\biggr\}\,,
\end{align}
and they all vanish in the limit $j=0$, so we find $\mathcal{Z}_{1}=0$ and then
\begin{equation}\label{E:dgetys}
	\mathcal{Z}_{V}=\mathcal{Z}_{}+\mathcal{O}(\lambda^{2})\,,
\end{equation}
that is, there is no first-order correction of $\mathcal{Z}_{V}$ from the nonlinear potential.

\subsection{Quantum statistical averages of operators in nonlinear systems}\label{S:ewbehrfwa}

The quantum statistical average of operator $\hat{\mathcal{O}}$ consisting of the system variables, as noted earlier, is given by
\begin{align}	
\langle\,\hat{\mathcal{O}}\,\rangle =\frac{1}{\mathcal{Z}_{V}}\,\operatorname{Tr}\Bigl\{\mathcal{O}\,\rho_{\chi}^{(V)}(t_{b})\Bigr\}
=\frac{1}{\mathcal{Z}_{V}}\int\!dq_{b}\,dr_{b}\;\delta(q_{b})\,\mathcal{O}(\chi_{b},p_{b})\,\rho_{\chi}(q_{b},r_{b},t)\,.
\end{align} 
In contrast to the generating functional, these expectation values may still have non-zero first-order nonlinear contributions. In particular, we are interested in the expectation values of the covariance matrix elements because they can be used to construct quantities we are interested in for the Gaussian system, or the nonGaussian system in terms of the perturbative treatment from the Gaussian system. Here we will use as an example the first-order correction of $\langle\hat{\chi}_{b}\rangle$ due to the nonlinear potential.

For $\langle\,\hat{\chi}_{b}\,\rangle$, we have its first-order nonlinear contribution $\langle\,\hat{\chi}_{b}\,\rangle^{(1)}$ given by
\begin{align}
	 \langle\,\hat{\chi}_{b}\,\rangle^{(1)}&=-i\,\frac{\lambda}{\mathcal{Z}_{V}}\int_{0}^{t}\!d\tau\;\biggl\{\frac{1}{2!}\left(\frac{\delta}{i\,\delta j_{q}(\tau)}\right)^{2}\left(\frac{\delta}{i\,\delta j_{r}(\tau)}\right)+\frac{1}{4!}\left(\frac{\delta}{i\,\delta j_{r}(\tau)}\right)^{3}\biggr\}\,\Xi[j;t)\mathcal{Z}[j;t)\;\bigg|_{j=0}\,,
\end{align}
due to \eqref{E:gkhers} and \eqref{E:dgetys}.

It is straightforward to show that
\begin{align}
	-i\,\frac{\lambda}{\mathcal{Z}_{V}}\int_{0}^{t}\!d\tau\;\biggl\{\frac{1}{4!}\left(\frac{\delta}{i\,\delta j_{r}(\tau)}\right)^{3}\biggr\}\,\Xi[j;t)\mathcal{Z}[j;t)\;\bigg|_{j_{q}=0=j_{r}}&=0\,,\notag
\intertext{but}
	-i\,\frac{\lambda}{\mathcal{Z}_{V}}\int_{0}^{t}\!d\tau\;\biggl\{\frac{1}{2!}\left(\frac{\delta}{i\,\delta j_{q}(\tau)}\right)^{2}\left(\frac{\delta}{i\,\delta j_{r}(\tau)}\right)\biggr\}\;\Xi[j;t)\mathcal{Z}[j;t)\;\bigg|_{j_{q}=0=j_{r}}&=-\frac{\lambda}{2}\int_{0}^{t}\!ds\;G_{R,0}^{(\chi)}(t-s)\,G_{H,0}^{(\chi)}(s,s)\,,\notag
\end{align}
for $0<s<t$. Therefore we have a nonvanishing  first-order nonlinear contribution to $\langle\hat{\chi}_{b}\rangle$ due to the anharmonic potential \eqref{E:ukbwzaf},
\begin{align}\label{E:eronksjaw}
	 \langle\,\hat{\chi}_{b}\,\rangle^{(1)}&=-\frac{\lambda}{2}\int_{0}^{t}\!ds\;G_{R,0}^{(\chi)}(t-s)\,G_{H,0}^{(\chi)}(s,s)\,.
\end{align}
{The cubic potential thus introduces a correction to the mean displacement from $\langle\,\hat{\chi}_{b}\,\rangle^{(0)}=0$. There is also a similar effect for a classical cubic oscillator. Their connection can be seen once we observe that  {$G_{H,0}^{(\chi)}(s,s)$ in the integrand in fact is proportional to} $\langle\,\hat{\chi}^{2}(s)\,\rangle^{(0)}$.}

Similarly we can show that $\langle\,\hat{p}_{b}\, \rangle$ also have nonvanishing first-order correction from the potential
\begin{align}
	\langle\,\hat{p}_{b}\,\rangle^{(1)}&=-\frac{m\lambda}{2!}\int_{0}^{t}\!ds\;\dot{G}_{R,0}^{(\chi)}(t-s)\,G_{H,0}^{(\chi)}(s,s)\,.
\end{align}
The expressions and their derivations are collected in Appendix~\ref{S:rtbsfgsre}. {It can also be simply inferred from the fact that $\hat{p}(\tau)=m\,\dot{\hat{\chi}}(\tau)$.}   The covariance matrix elements are useful for exploring many interesting physical problems, such as the approach to equilibrium and the existence of a fluctuation-dissipation relation at late times  in weakly anharmonic oscillator open quantum systems~\cite{NLFDR}, or energy transport and the existence of nonequilibrium steady state in weakly nonlinear quantum systems with multiple baths, based on the results of the linear cases investigated before (e.g.,~\cite{HPZ92,HHAoP}).  These problems are investigated in the present series of papers.


\subsection{Real-time two-point functions of the anharmonic oscillator}\label{S:dhtebe}
We will find various two-point functions of the nonlinear oscillator in terms of the functional method. Here we consider the nonlinear potential of the form
\begin{align}
	V(\chi)&=\frac{\lambda}{4!}\,\chi^{4}\,,\qquad\qquad&\lambda&>0
\end{align}
instead of the $\chi^{3}$ theory, as an additional application. The corresponding first-order correction of the generating functional due to the nonlinear potential is then given by
\begin{align}
	\mathcal{Z}_{1}[j]&=-i\lambda\int_{0}^{t}\!ds\;\biggl\{\frac{1}{3!}\frac{\delta^{4}}{\delta j_{q}^{3}(s)\delta j_{r}(s)}+\frac{1}{4!}\frac{\delta^{4}}{\delta j_{q}(s)\delta j_{r}^{3}(s)}\biggr\}\,\mathcal{Z}[j]\notag\\
		&=-i\lambda\int_{0}^{t}\!ds\;\biggl\{\frac{1}{2!}\,\mathfrak{J}_{q}(s)G_{H,0}^{(\chi)}(s,s)\Xi(s)\mathcal{Z}+\frac{1}{3!}\,\mathfrak{J}_{q}(s)\Xi^{3}(s)\mathcal{Z}+\frac{1}{4!}\,\mathfrak{J}_{q}^{3}(s)\Xi(s)\mathcal{Z}\biggr\}\,.\label{E:hgvjrts}
\end{align}
in which we have used
\begin{align}
	\frac{\delta^{4}\mathcal{Z}}{\delta j_{q}^{3}(s)\delta j_{r}(s)}&=3i^{4}\mathfrak{J}_{q}(s)G_{H,0}^{(\chi)}(s,s)\Xi(s)\mathcal{Z}+i^{4}\mathfrak{J}_{q}(s)\Xi^{3}(s)\mathcal{Z}\,,&\frac{\delta^{4}\mathcal{Z}}{\delta j_{q}(s)\delta j_{r}^{3}(s)}&=i^{4}\mathfrak{J}_{q}^{3}(s)\Xi(s)\mathcal{Z}\,,
\end{align}
Since the Feynman propagator of the nonlinear oscillator can be calculated according to \eqref{E:tbdfse},
\begin{align*}
	\langle\,\mathcal{T}\,\hat{\chi}(\tau)\hat{\chi}(\tau')\,\rangle&=-\frac{1}{\mathcal{Z}_{V}[j;t)}\left\{\frac{\delta^{2}\mathcal{Z}_{V}[j;t)}{\delta j_{q}(\tau)\,\delta j_{q}(\tau')}+\frac{1}{2}\frac{\delta^{2}\mathcal{Z}_{V}[j;t)}{\delta j_{q}(\tau')\delta j_{r}(\tau)}+\frac{1}{2}\frac{\delta^{2}\mathcal{Z}_{V}[j;t)}{\delta j_{q}(\tau)\,\delta j_{r}(\tau')}+\frac{1}{4}\frac{\delta^{2}\mathcal{Z}_{V}[j;t)}{\delta j_{r}(\tau)\,\delta j_{r}(\tau')}\right\}\;\bigg|_{\substack{j=0\\q=0}}\,,
\end{align*}
where $\mathcal{Z}_{V}=\mathcal{Z}+\mathcal{Z}_{1}+\mathcal{O}(\lambda^{2})$ is the generating functional of the nonlinear oscillator, we find
\begin{align}
	\langle\mathcal{T}\hat{\chi}(\tau)\hat{\chi}(\tau')\rangle^{(1)}&=\lambda\int_{0}^{t}\!ds\,\biggl\{-\frac{1}{2}\,G_{R,0}^{(\chi)}(\tau-s)G_{H,0}^{(\chi)}(s,s)G_{H,0}^{(\chi)}(s,\tau')-\frac{1}{2}\,G_{R,0}^{(\chi)}(\tau'-s)G_{H,0}^{(\chi)}(s,s)G_{H,0}^{(\chi)}(s,\tau)\biggr.\notag\\
	&\qquad\qquad\;+\biggl.\frac{i}{4}\biggl[G_{R,0}^{(\chi)}(\tau-s)G_{H,0}^{(\chi)}(s,s)G_{R,0}^{(\chi)}(s-\tau')+G_{R,0}^{(\chi)}(\tau'-s)G_{H,0}^{(\chi)}(s,s)G_{R,0}^{(\chi)}(s-\tau)\biggr]\biggr\}\,,
\end{align}
with the help of \eqref{E:bgfhke1} and \eqref{E:bgfhke2}.  In Sec.~\ref{S:rbkher}, for the zeroth order case, we have made such identifications
\begin{align}
	\langle\mathcal{T}\hat{\chi}(\tau)\hat{\chi}(\tau')\rangle^{(0)}&=G_{H,0}^{(\chi)}(\tau,\tau')-\frac{i}{2}\,G_{A,0}^{(\chi)}(\tau,\tau')-\frac{i}{2}\,G_{R,0}^{(\chi)}(\tau,\tau')\,.
\end{align}
This implies that
\begin{align}
	G_{H,1}^{(\chi)}(\tau,\tau')&=-\frac{\lambda}{2}\int_{0}^{t}\!ds\;\Bigl[G_{R,0}^{(\chi)}(\tau-s)G_{H,0}^{(\chi)}(s,s)G_{H,0}^{(\chi)}(s,\tau')+G_{R,0}^{(\chi)}(\tau'-s)G_{H,0}^{(\chi)}(s,s)G_{H,0}^{(\chi)}(s,\tau)\Bigr]\,,\label{E:oerwpijen1}\\
	G_{R,1}^{(\chi)}(\tau,\tau')&=-\frac{\lambda}{2}\int_{0}^{t}\!ds\;G_{R,0}^{(\chi)}(\tau-s)G_{H,0}^{(\chi)}(s,s)G_{R,0}^{(\chi)}(s-\tau')\,,\label{E:oerwpijen2}
\end{align}
with the assumption $0<\tau'\leq\tau<t$.  {We have used the following notation for the Green's function: the superscript indicates the subsystem it is associated with, the subscript contains information about the type of Green's function and its perturbative order. For example,   $G_{H,1}^{(\chi)}$ denotes the first-order correction of the oscillator's Hadamard function for the anharmonic potential.} From \eqref{E:oerwpijen1} and \eqref{E:oerwpijen2}, we see in general the first-order correction is not stationary even though one of their zeroth-order counterparts are.  {This inevitably introduces complexity in the  analysis of the dynamical evolution of the correction terms.}

To show that the FDR is satisfied for the first-order correction due to the nonlinear potential, we need to first address the nonstationarity issues in \eqref{E:oerwpijen1} and \eqref{E:oerwpijen2}. It is easier to start with the first-order correction of the retarded function $G_{R,1}^{(\chi)}$ of the oscillator. At late times $\tau$, $\tau'\gg\gamma^{-1}$, the Hadamard function $H_{H,0}^{(\chi)}(\tau,\tau')$ of the free oscillator, that is, the zeroth-order contribution, becomes stationary~\cite{QTD,HHL18}, so we can write \eqref{E:oerwpijen2} as
\begin{align}
	G_{R,1}^{(\chi)}(\tau,\tau')&\simeq-\frac{\lambda}{2}\int_{\tau'}^{\tau}\!ds\;G_{R,0}^{(\chi)}(\tau-s)G_{H,0}^{(\chi)}(s-s)G_{R,0}^{(\chi)}(s-\tau')\notag\\
	&=-\frac{\lambda}{2}\,G_{H,0}^{(\chi)}(0)\int_{-\infty}^{\infty}\!\frac{d\kappa}{2\pi}\;\bigl[\widetilde{G}_{R,0}^{(\chi)}(\kappa)\Bigr]^{2}e^{-i\kappa(\tau-\tau')}\,.
\end{align}
Thus we see it indeed becomes stationary at late times. It is trickier to show stationarity for $G_{H,1}^{(\chi)}(\tau,\tau')$. In general, the Hadamard function of the free oscillator is given by
\begin{align}
	G_{H,0}^{(\chi)}(\tau,\tau')&=D_{1}(\tau)D_{1}(\tau')\,\langle\hat{\chi}^{2}(0)\rangle+D_{2}(\tau)D_{2}(\tau')\,\langle\dot{\hat{\chi}}^{2}(0)\rangle\notag\\
	&\qquad\qquad\qquad\qquad\qquad\qquad+\frac{e^{2}}{m^{2}}\int_{-\infty}^{\infty}\frac{d\kappa}{2\pi}\;\widetilde{G}_{H,0}^{(\phi)}(\mathbf{0};\kappa)\,e^{-i\kappa(\tau-\tau')}f_{\kappa}(\tau)f_{\kappa}^{*}(\tau')\,,
\end{align}
with
\begin{align}
	f_{\kappa}(s)=\int_{0}^{s}\!dy\;D_{2}(y)\,e^{+i\kappa y}&=\widetilde{D}_{2}(\kappa)+e^{-\gamma s}\times(\text{sinusoidal functions of $s$})\,.\label{E:roiadx}
\end{align}
The second term in \eqref{E:roiadx}, as well as $D_{1}(s)$ and $D_{2}(s)$, will be exponentially small in the limit $s\gg\gamma^{-1}$. This implies the expression, say, $G_{H,0}^{(\chi)}(s,\tau')$ in \eqref{E:oerwpijen1} will be approximately given by
\begin{align}
	G_{H,0}^{(\chi)}(s,\tau')&\simeq\frac{e^{2}}{m^{2}}\int_{-\infty}^{\infty}\frac{d\kappa}{2\pi}\;\widetilde{G}_{H,0}^{(\phi)}(\mathbf{0};\kappa)f_{\kappa}(s)f_{\kappa}^{*}(\tau')\label{E:firthsw}\\
		&= e^{2}\!\int_{-\infty}^{\infty}\frac{d\kappa}{2\pi}\;\biggl[\widetilde{G}_{H,0}^{(\phi)}(\mathbf{0};\kappa)\widetilde{G}_{R,0}^{(\chi)*}(\kappa)\widetilde{G}_{R,0}^{(\chi)}(\kappa)\,e^{-i\kappa(s-\tau')}+e^{-\gamma s}\times(\text{sinusoidal functions of $s$})\biggr]\,,\notag
\end{align}
for $\tau'\gg\gamma^{-1}$. On the other hand, we observe that $G_{H,0}^{(\chi)}(s,s)$ also takes a form similar to \eqref{E:roiadx}, except that the part proportional to  $e^{-\gamma s}$ may not be exponentially small, and it remains bounded if we introduce a cutoff scale to the evaluation of the expression.

With these in mind, we rewrite one of the integrals in \eqref{E:oerwpijen1} by
\begin{align}
	&\quad\int_{0}^{\tau}\!ds\;G_{R,0}^{(\chi)}(\tau-s)G_{H,0}^{(\chi)}(s,s)G_{H,0}^{(\chi)}(s,\tau')\notag\\
	&=\int_{-\infty}^{\infty}\!\frac{d\kappa}{2\pi}\;\widetilde{G}_{H,0}^{(\phi)}(\kappa)\,e^{-i\kappa(\tau-\tau')}\int_{0}^{\tau}\!ds\;G_{R,0}^{(\chi)}(s)G_{H,0}^{(\chi)}(\tau-s,\tau-s)e^{+i\kappa s}f_{\kappa}(\tau-s)f_{\kappa}^{*}(\tau')\,,\label{E:fghrhrwsf}
\end{align}
according to \eqref{E:firthsw} for $0<\gamma^{-1}\ll\tau,\,\tau'<t$. We now evaluate the integral over $s$ in \eqref{E:fghrhrwsf}. Note that $G_{R,0}^{(\chi)}(s)\propto e^{-\gamma s}$, and $f(\tau-s)\propto \text{const.}+e^{-\gamma(\tau-s)}\times\cdots$, that is, explicitly,
\begin{align}
	G_{R,0}^{(\chi)}(s)&= e^{-\gamma s}\times\bigl(\cdots\bigr)\,,\\
	G_{H,0}^{(\chi)}(\tau-s,\tau-s)&= e^{-2\gamma(\tau-s)}\times\bigl(\cdots\bigr)+\frac{e^{2}}{m^{2}}\int_{-\infty}^{\infty}\frac{d\kappa}{2\pi}\;\widetilde{G}_{H,0}^{(\phi)}(\mathbf{0};\kappa)\,\lvert\text{const.}+e^{-\gamma(\tau-s)}\times\bigl(\cdots\bigr)\rvert^{2}\,,\label{E:fgkrsder}\\
	f_{\kappa}(\tau-s)&= \text{const.}+e^{-\gamma(\tau-s)}\times\bigl(\cdots\bigr)
\end{align}
where $(\cdots)$ represents terms that are sinusoidal in $s$, so we are generically dealing with integrals of the forms
\begin{align}
	\int_{0}^{\tau}\!ds\;e^{-\gamma s}&=\text{const.}+e^{-\gamma\tau}\times\bigl(\cdots\bigr)\,,&\int_{0}^{\tau}\!ds\;e^{-\gamma s}e^{+i\varpi s}&=e^{-\gamma\tau}\times\bigl(\cdots\bigr)\,,\label{E:gfkrt1}\\
	&&\int_{0}^{\tau}\!ds\;e^{-\gamma s}e^{-n\gamma (\tau-s)}e^{+i\varpi s}&=e^{-n\gamma\tau}\times\bigl(\cdots\bigr)\,.\label{E:gfkrt2}
\end{align}
for $n=1$, 2, 3 and $\varpi\in\mathbb{R}$. Therefore we see the integral over $s$ in \eqref{E:fghrhrwsf} will give a result that depends on $\kappa$ but is independent of $\tau$ for sufficiently large $\tau$ and $\tau'$. In turn, it implies that \eqref{E:fghrhrwsf} is stationary, that is, it is a function of $\tau-\tau'$. Therefore an important conclusion for a $\chi^{4}$ anharmonic oscillator coupled to a quantum field bath  is that its first-order corrections of the retarded Green's function and the Hadamard function  will become  stationary at late times, as their zeroth-order counterparts. We then have
\begin{align}
	G_{H,1}^{(\chi)}(\tau,\tau')&=G_{H,1}^{(\chi)}(\tau-\tau')\,,&G_{R,1}^{(\chi)}(\tau,\tau')&=G_{R,1}^{(\chi)}(\tau-\tau')\,,
\end{align}
for $\gamma^{-1}\ll\tau,\,\tau'$.

Now we note that at late time, the remaining nontrivial contribution of the $s$ integral in \eqref{E:fghrhrwsf} comes from the $s$-independent components of $G_{H,0}^{(\chi)}(\tau-s,\tau-s)$ and $G_{H,0}^{(\chi)}(s,\tau')$, so that this nontrivial contribution is given by
\begin{align}
	\int_{0}^{\tau}\!ds\;G_{R,0}^{(\chi)}(\tau-s)G_{H,0}^{(\chi)}(s,s)G_{H,0}^{(\chi)}(s,\tau')&\simeq\langle\hat{\chi}^{2}(\infty)\rangle\int_{-\infty}^{\infty}\!\frac{d\kappa}{2\pi}\;\widetilde{G}_{H,0}^{(\chi)}(\kappa)\widetilde{G}_{R,0}^{(\chi)}(\kappa)\,e^{-i\kappa(\tau-\tau')}\,,
\end{align}
where we have used the result for the free harmonic oscillator coupled to the scalar field~\cite{QTD,HHL18}
\begin{align}
	\lim_{\tau,\tau'\to\infty}G_{H,0}^{(\chi)}(\tau,\tau')=G_{H,0}^{(\chi)}(\tau-\tau')&=\int\!\frac{d\kappa}{2\pi}\;\coth\frac{\beta\kappa}{2}\,\operatorname{Im}\Bigl\{\widetilde{G}_{R,0}^{(\chi)}(\kappa)\Bigr\}\,e^{-i\,\kappa(\tau-\tau')}\,,\label{E:euhfshf}
\end{align}
and the FDR of the free scalar field
\begin{equation}
	\widetilde{G}_{H,0}^{(\phi)}(\kappa)=\coth\frac{\beta\kappa}{2}\,\operatorname{Im}\widetilde{G}_{R,0}^{(\phi)}(\kappa)\,.
\end{equation}
Therefore we obtain from \eqref{E:oerwpijen1} that
\begin{align}
	G_{H,1}^{(\chi)}(\tau-\tau')&=-\lambda\,\langle\hat{\chi}^{2}(\infty)\rangle\int_{-\infty}^{\infty}\!\frac{d\kappa}{2\pi}\;\widetilde{G}_{H,0}^{(\chi)}(\kappa)\,\operatorname{Re}\bigl\{\widetilde{G}_{R,0}^{(\chi)}(\kappa)\bigr\}\,e^{-i\kappa(\tau-\tau')}\,,
\end{align}
at late times. As a consequence we find
\begin{align}
	\widetilde{G}_{H,1}^{(\chi)}(\kappa)&=-\lambda\,\langle\hat{\chi}^{2}(\infty)\rangle\,\coth\frac{\beta\kappa}{2}\,\operatorname{Im}\bigl\{\widetilde{G}_{R,0}^{(\chi)}(\kappa)\bigr\}\operatorname{Re}\bigl\{\widetilde{G}_{R,0}^{(\chi)}(\kappa)\bigr\}\,,\label{E:ribsierb1}
\end{align}
at late times.

As for the first-order correction of the retarded Green's function of the nonlinear oscillator, we can write \eqref{E:oerwpijen2} as
\begin{align}
	G_{R,1}^{(\chi)}(\tau,\tau')&=-\frac{\lambda}{2}\int_{\tau'}^{\tau}\!ds\;G_{R,0}^{(\chi)}(\tau-s)G_{H,0}^{(\chi)}(s,s)G_{R,0}^{(\chi)}(s-\tau')\notag\\
	&\simeq-\frac{\lambda}{2}\,\langle\hat{\chi}^{2}(\infty)\rangle\int_{-\infty}^{\infty}\!\frac{d\kappa}{2\pi}\;\,\widetilde{G}_{R,0}^{(\chi)}(\kappa)\widetilde{G}_{R,0}^{(\chi)}(\kappa)\,e^{-i\kappa(\tau-\tau')}\,,\label{E:gbheeas}
\end{align}
{by directly evaluating the $s$ integral,} for $0<\gamma^{-1}<\tau,\,\tau'<t$. Note that \eqref{E:oerwpijen2} or \eqref{E:gbheeas} has implied the presence of $\theta(\tau-\tau')$. From Eq.~\eqref{E:gbheeas}, we can read off the Fourier transform of $G_{R,1}^{(\chi)}(\tau,\tau')$ as
\begin{align}
	\widetilde{G}_{R,1}^{(\chi)}(\kappa)=-\frac{\lambda}{2}\,\langle\hat{\chi}^{2}(\infty)\rangle\,\Bigl[\widetilde{G}_{R,0}^{(\chi)}(\kappa)\Bigr]{}^{2}\,,
\end{align}
so its imaginary part is given by
\begin{align}
	\operatorname{Im}\widetilde{G}_{R,1}^{(\chi)}(\kappa)&=-\lambda\,\langle\hat{\chi}^{2}(\infty)\rangle\,\operatorname{Im}\bigl\{\widetilde{G}_{R,0}^{(\chi)}(\kappa)\bigr\}\operatorname{Re}\bigl\{\widetilde{G}_{R,0}^{(\chi)}(\kappa)\bigr\}\,.\label{E:ribsierb2}
\end{align}
Comparing \eqref{E:ribsierb1} and \eqref{E:ribsierb2}, we arrive at the FDR of the nonlinear oscillator perturbatively to the first order in the self-coupling constant $\lambda$
\begin{equation}\label{E:rturtdfjw}
	\widetilde{G}_{H,1}^{(\chi)}(\kappa)=\coth\frac{\beta\kappa}{2}\,\operatorname{Im}\widetilde{G}_{R,1}^{(\chi)}(\kappa)\,.
\end{equation}
It is interesting to stress that in the derivation of \eqref{E:rturtdfjw}, although we have taken advantage of {the anharmonicity being small, we need not assume weak coupling between the oscillator and the quantum field bath. Thus the evolution of the anharmonic oscillator is fully dynamical  (in contrast to linear response theory) and its final equilibrium state after the nonequilibrium evolution, if it exits, can markedly be different from  its initial state}. In addition, we observe that two prerequisite conditions play  important roles in these perturbative arguments. The first condition is the existence of the equilibrium state for the zeroth order (linear or harmonic) case. This equilibrium state has the characteristics that it is approached exponentially and it is independent of the initial states of the oscillator. This warrants the possibility of an equilibrium state for a nonequilibrium weakly nonlinear oscillator when it is coupled to a environmental quantum field {as long as the perturbative expansion is legitimate}. The second, but accompanying, condition lies in the fact that the nonstationary component of the two-point functions of the weakly nonlinear oscillator vanishes exponentially at the time scale greater than the relaxation time scale, introduced in the zeroth order dynamics.

These two conditions {need to be satisfied in our nonperturbative derivation of the FDR~\cite{NLFDR}, namely,} the existence of the equilibrium state and the subdominance of the nonstationary components of the correlation functions.  {This will be discussed in our follow-up paper~\cite{NLFDR}. So far, we have shown that after the anharmonic oscillator undergoes   nonequilibrium evolution from its initial state, its two-point functions indeed become stationary at late times, and an FDR emerges. We can further ask if the emergence of an FDR is directly linked to the existence of the final equilibrium state? In Sections~\ref{S:fieryga} and~\ref{S:ertbfkjd}, we first provide a general argument, borrowed from~\cite{NLFDR}, and then give a detailed perturbative calculation of the energy exchange between the oscillator and the quantum field bath to provide an affirmative answer to this question.}

{\subsection{The Four-point Function}
In this subsection we provide a more complicated example to show the advantage of the functional perturbative method. Consider one type of four-point function and its first-order correction for the $\chi^{4}$ anharmonic oscillator coupled to the quantum field bath. These are used in the calculation of the power delivered by the anharmonic  potential  and the higher-order contributions.

 We first  write \eqref{E:dbeirdsf} explicitly as
\begin{align}\label{E:lppwesjser}
	\mathcal{Z}[j_{\pm}]&=\exp\biggl\{\frac{i}{2}\int_{0}^{t}\!ds\,ds'\;\biggl[j_{+}(s)\,G_{++,0}^{(\chi)}(s,s')\,j_{+}(s')-j_{+}(s)\,G_{+-,0}^{(\chi)}(s,s')\,j_{-}(s')\Bigr.\biggr.\\
	&\qquad\qquad\qquad\qquad\qquad\qquad\qquad-\biggl.\Bigl.j_{-}(s)\,G_{-+,0}^{(\chi)}(s,s')\,j_{+}(s')+j_{-}(s)\,G_{--,0}^{(\chi)}(s,s')\,j_{-}(s')\biggr]\biggr\}\,.\notag
\end{align}
It   generates the following useful expressions
\begin{align}
	\frac{\delta\mathcal{Z}}{\delta j_{+}(\tau)}&=i\,\Theta(\tau)\,\mathcal{Z}&\frac{\delta\mathcal{Z}}{\delta j_{-}(\tau)}&=i\,\Upsilon(\tau)\,\mathcal{Z}\,,
\end{align}
where
\begin{align}
	\Theta(\tau)&=\frac{1}{2}\int_{0}^{t}\!ds\;\Bigl\{+\Bigl[G_{++,0}^{(\chi)}(\tau,s)+G_{++,0}^{(\chi)}(s,\tau)\Bigr]j_{+}(s)-\Bigl[G_{+-,0}^{(\chi)}(\tau,s)+G_{-+,0}^{(\chi)}(s,\tau)\Bigr]j_{-}(s)\Bigr\}\,,\\
	\Upsilon(\tau)&=\frac{1}{2}\int_{0}^{t}\!ds\;\Bigl\{-\Bigl[G_{-+,0}^{(\chi)}(\tau,s)+G_{+-,0}^{(\chi)}(s,\tau)\Bigr]j_{+}(s)+\Bigl[G_{--,0}^{(\chi)}(\tau,s)+G_{--,0}^{(\chi)}(s,\tau)\Bigr]j_{-}(s)\Bigr\}\,,
\end{align}
such that
\begin{align}
	\frac{\delta^{2}\mathcal{Z}}{\delta j_{+}(\tau)j_{+}(\tau')}&=+i\,G_{++,0}^{(\chi)}(\tau,\tau')\,\mathcal{Z}-\Theta(\tau)\Theta(\tau')\,\mathcal{Z}\,,\label{E:tvsdf1}\\
	\frac{\delta^{2}\mathcal{Z}}{\delta j_{+}(\tau)j_{-}(\tau')}&=-i\,G_{+-,0}^{(\chi)}(\tau,\tau')\,\mathcal{Z}-\Theta(\tau)\Upsilon(\tau')\,\mathcal{Z}\,,\\
	\frac{\delta^{2}\mathcal{Z}}{\delta j_{-}(\tau)j_{+}(\tau')}&=-i\,G_{-+,0}^{(\chi)}(\tau,\tau')\,\mathcal{Z}-\Upsilon(\tau)\Theta(\tau')\,\mathcal{Z}\,,\\
	\frac{\delta^{2}\mathcal{Z}}{\delta j_{-}(\tau)j_{-}(\tau')}&=+i\,G_{--,0}^{(\chi)}(\tau,\tau')\,\mathcal{Z}-\Upsilon(\tau)\Upsilon(\tau')\,\mathcal{Z}\,.\label{E:tvsdf4}
\end{align}
By taking $j\to0$ limit, we can relate $G_{\pm\pm,0}^{(\chi)}(\tau,\tau')$ in \eqref{E:tvsdf1}--\eqref{E:tvsdf4} with the path-ordered two-point functions \eqref{E:fkjtbfdgf} of the oscillator.}

 {In the limit $j_{\pm}\to0$, we have
\begin{equation}\label{E:fghesczx}
	\langle\mathcal{T}\hat{\chi}^{3}(\tau)\hat{\chi}(\tau')\rangle^{(0)}=\frac{1}{\mathcal{Z}}\frac{\delta^{4}\mathcal{Z}}{\delta j_{+}^{3}(\tau)\delta j_{+}(\tau')}\,\bigg|_{j=0}=-3G_{++,0}^{(\chi)}(\tau,\tau)G_{++,0}^{(\chi)}(\tau,\tau')=-3G_{F,0}^{(\chi)}(\tau,\tau)G_{F,0}^{(\chi)}(\tau,\tau')\,.
\end{equation}
For the first-order correction of \eqref{E:fghesczx}, we need to calculate 
\begin{equation}
	-i\,\frac{\lambda}{4!\mathcal{Z}}\int_{0}^{t}\!ds\biggl[\frac{\delta^{4}}{\delta j_{+}^{4}(s)}-\frac{\delta^{4}}{\delta j_{-}^{4}(s)}\biggr]\frac{\delta^{4}\mathcal{Z}}{\delta j_{+}^{3}(\tau)\delta j_{+}(\tau')}\,\bigg|_{j=0}\,.
\end{equation}
We find
\begin{align}
	&\quad\frac{1}{\mathcal{Z}}\biggl[\frac{\delta^{4}}{\delta j_{+}^{4}(s)}-\frac{\delta^{4}}{\delta j_{-}^{4}(s)}\biggr]\frac{\delta^{4}\mathcal{Z}}{\delta j_{+}^{3}(\tau)\delta j_{+}(\tau')}\,\bigg|_{j=0}\notag\\
	&=i\,36G_{++,0}^{(\chi)}(\tau,\tau)\Bigl[G_{++,0}^{(\chi)}(\tau,s)G_{++,0}^{(\chi)}(\tau',s)-G_{+-,0}^{(\chi)}(\tau,s)G_{+-,0}^{(\chi)}(\tau',s)\Bigr]G_{H,0}^{(\chi)}(s,s)\notag\\
	&\qquad\qquad\qquad+i\,36G_{++,0}^{(\chi)}(\tau,\tau')\Bigl[G_{++,0}^{(\chi)2}(\tau,s)-G_{+-,0}^{(\chi)2}(\tau,s)\Bigr]G_{H,0}^{(\chi)}(s,s)\notag\\
	&\qquad\qquad\qquad\qquad\qquad\qquad+24\Bigl[G_{++,0}^{(\chi)3}(\tau,s)G_{++,0}^{(\chi)}(\tau',s)-G_{+-,0}^{(\chi)3}(\tau,s)G_{+-,0}^{(\chi)}(\tau',s)\Bigr]\,.\label{E:otubdjfsa}
\end{align}
Here we explicitly see the trend that the higher-order terms produce the products of the mixed types of the zeroth-order Green's functions such as $G_{++,0}^{(\chi)}$ and $G_{+-,0}^{(\chi)}$ in this case. It may not be so obvious in the operator approach in deriving the same result.}

 {From \eqref{E:otubdjfsa}, we then obtain 
\begin{align}\label{E:rbudfs}
	\langle\mathcal{T}\hat{\chi}^{3}(\tau)\hat{\chi}(\tau')\rangle^{(1)}&=-i\,\frac{\lambda}{4!}\int_{0}^{t}\!ds\;\eqref{E:otubdjfsa}\notag\\
	&=-3\Bigl[G_{++,0}^{(\chi)}(\tau,\tau)G_{++,1}^{(\chi)}(\tau,\tau')+G_{++,0}^{(\chi)}(\tau,\tau')G_{++,1}^{(\chi)}(\tau,\tau)\Bigr]\\
	&\qquad\qquad\qquad\qquad-i\,\lambda\int_{0}^{t}\!ds\;\Bigl[G_{++,0}^{(\chi)3}(\tau,s)G_{++,0}^{(\chi)}(\tau',s)-G_{+-,0}^{(\chi)3}(\tau,s)G_{+-,0}^{(\chi)}(\tau',s)\Bigr]\,.\notag
\end{align}
We see the disconnected component, which is an obvious extension of \eqref{E:fghesczx} to the higher-order. In addition, there is the connected component, written as a convolution integral. These are similar to the contributions of common higher-loop Feynman diagrams in nonequilibrium field theory~\cite{Jo86,CalHu2008}. Notice an interesting arrangement of the contributions from different time branches.}

 {We may equally well express \eqref{E:rbudfs} in terms of $G_{R,0}^{(\chi)}$ and $G_{F,0}^{(\chi)}$ we used earlier by the identities valid for general two-point functions,
\begin{align}
	G_{F}(\tau,\tau')&=G_{++}(\tau,\tau')=i\,G_{H}(\tau,\tau')+\frac{1}{2}\,G_{R}(\tau,\tau')+\frac{1}{2}\,G_{R}(\tau',\tau)\,,\label{E:dgnek1}\\
	G_{S}(\tau,\tau')&=G_{-+}(\tau,\tau')=i\,G_{H}(\tau,\tau')+\frac{1}{2}\,G_{R}(\tau,\tau')-\frac{1}{2}\,G_{R}(\tau',\tau)=G_{>}(\tau,\tau')\,,\\
	G_{S}(\tau',\tau)&=G_{+-}(\tau,\tau')=i\,G_{H}(\tau,\tau')-\frac{1}{2}\,G_{R}(\tau,\tau')+\frac{1}{2}\,G_{R}(\tau',\tau)=G_{<}(\tau,\tau')\,.\label{E:dgnek3}
\end{align}
However, the result is somewhat cumbersome. In the next section, we shall address the physical meaning of the FDR, and its connection with the dynamical equilibration of the anharmonic oscillator coupled to the quantum field bath, in terms of the their mutual exchange of energy.}

\section{Energy Balance between an Anharmonic Oscillator and the Quantum Field}\label{S:fieryga}

{When a weakly anharmonic oscillator, which can be taken as representing the internal degrees of freedom of an Unruh-DeWitt detector, is coupled to a quantum field environment, the interaction term governs the energy exchange between the oscillator and the field. If there exists an equilibrium state for the dynamics of the reduced system, say, the anharmonic oscillator, then we expect the energy exchange between the system and its environment should come to balance. Thus checking on how this net energy flow varies in time and observing when it ceases to exist will provide us information about the relaxation process of the reduced system.}

{The formalism developed earlier already enables us to compute the expectation values of the physical observables associated with the anharmonic oscillator, and thus  allows us to follow their time evolutions, at least in a perturbative sense. It offers a direct route to examine the existence of the relaxed equilibrium state in the anharmonic oscillator's dynamics. Here we will present a two stage progression. First, to paint a more transparent physical picture, we will use the quantum Langevin equation of the anharmonic oscillator to illustrate the dynamics, to identify the physical quantities that are relevant in the context of energy exchange between the oscillator and the bath field, and discuss the corresponding implications. Second, after that, we will use the functional method  developed earlier  to compute these quantities and their leading order correction due to anharmonicity. The motivation for this dual approach comes from the observations that 1) the functional method is particularly powerful in revealing the underlying structure of the higher-order contributions in the perturbation theory in terms of the zero-order two point functions of the system of interest. In other words,  we find it relatively straightforward to carry out by the functional method  the Wick-like expansion of the product of the oscillator's variables for a general Gaussian state than the canonical operator approach. Secondly, in  addition to the Feynman propagator, a relevant two-point function for  the in-out formalism, we encounter {in the in-in formalism} three other types of two-point functions, i.e., the retarded/advanced Green's function and the Dyson function (or their linear combinations by the Keldysh rotation). The functional method can  generate these Green's functions naturally, as seen earlier in our perturbative calculations. By comparison  the canonical operator approach seems less convenient for this task, although in principle we believe it can be achieved.}

{Here we consider a Unruh-DeWitt detector, whose internal degree of freedom is modeled by a quantum anharmonic oscillator, our system, coupled to an environment, described by a massless quantum scalar field $\hat{\phi}$ initially in a thermal state. Suppose the operator $\hat{\chi}$ accounts for the displacement of the oscillator 
and $\mathbf{z}$ denotes the external trajectory of the detector (It is not a dynamical variable in the current configuration, so it will not partake of the interaction between the internal degree of freedom of the detector and the bath field.) We can write down a} simultaneous set of Heisenberg equations for these two dynamical variables
\begin{align}
	\ddot{\hat{\chi}}(t)+\omega_{0}^{2}\hat{\chi}(t)+\lambda\,V'[\hat{\chi}(t)]&=\frac{e}{m}\,\hat{\phi}(\mathbf{z},t)\,,\label{E:vher1}\\
	\bigl(\partial_{t}^{2}-\pmb{\nabla}_{x}^{2}\bigr)\hat{\phi}(\mathbf{x},t)&=e\,\hat{\chi}(t)\,\delta^{(3)}(\mathbf{x}-\mathbf{z})\,,\label{E:vher2}
\end{align}
{where $m$ is the mass of the oscillator, $\omega_{0}$ is the bare frequency of the oscillator in the absence of the anharmonic potential $V$, whose self-coupling strength is given by $\lambda$, and $e$ measures the coupling between the detector $\hat{\chi}$ and the field $\hat{\phi}$. We can then} derive a reduced Heisenberg equation for the anharmonic oscillator
\begin{align}\label{E:djsh}
	\ddot{\hat{\chi}}(t)+\omega_{0}^{2}\hat{\chi}(t)+\lambda\,V'[\hat{\chi}(t)]&=\frac{e}{m}\,\hat{\phi}(\mathbf{z},t)=\frac{e}{m}\,\hat{\phi}_{h}(\mathbf{x},t)+\frac{e^{2}}{m}\int_{0}^{t}\!ds\;G_{R,\,0}^{(\phi)}(\mathbf{z},t;\mathbf{z},s)\,\hat{\chi}(s)\,.
\end{align}
We see that the oscillator is driven by a noise of the form $e\hat{\phi}_{h}(\mathbf{x},t)$ associated with the quantum fluctuations of the free scalar field $\hat{\phi}_{h}$, the homogeneous solution of the field equation \eqref{E:vher2}. The nonlocal expression on the righthand side of \eqref{E:djsh} describes the backreaction due to the radiation of the scalar field emitted from the oscillator, which constitutes a damping mechanism.  {This is most easily seen if we note that for a scalar field bath this nonlocal expression will reduce to a frequency renormalization and a damping term such that \eqref{E:djsh} becomes
\begin{equation}\label{E:gbkftes}
	\ddot{\hat{\chi}}(t)+2\gamma\,\dot{\hat{\chi}}(t)+\omega^{2}\hat{\chi}(t)+\lambda\,V'[\hat{\chi}(t)]=\frac{e}{m}\,\hat{\phi}_{h}(\mathbf{x},t)\,,
\end{equation}
with $\gamma=e^{2}/8\pi m$ being the damping constant and $\omega$ the renormalized frequency. Thus physically, \eqref{E:gbkftes} tells that the fluctuating force due to the free-field fluctuations will pump energy into the oscillator from the field, whereas the damping will disperse the energy of the oscillator back into its surrounding field~\cite{AFM0}. Both together will account for the energy exchange between the oscillator and the scalar field. In general, the net energy flow resulting from this exchange varies with time, depending on the states of both subsystems. However, if it asymptotically relaxes to a vanishing value, then it signifies the existence of an equilibrium state for the oscillator's dynamics.}

From \eqref{E:djsh}, it is natural to identify the power $P_{\xi}$ delivered by the noise force {$e\,\hat{\phi}_{h}(\mathbf{x},t)$ as}
\begin{equation}
	P_{\xi}(\tau)=\frac{e}{2}\,\langle\bigl\{\hat{\phi}_{h}(\mathbf{z},\tau),\,\dot{\hat{\chi}}(\tau)\bigr\}\rangle\,,
\end{equation}
and the power delivered by the backaction of radiation $P_{\gamma}$,
\begin{align}
	P_{\gamma}(\tau)&=\frac{e^{2}}{2}\int_{0}^{\tau}\!ds\;G_{R,\,0}^{(\phi)}(\mathbf{z},t;\mathbf{z},s)\langle\bigl\{\hat{\chi}(s),\,\dot{\hat{\chi}}(\tau)\bigr\}\rangle+\cdots\,,
\end{align}
where $\cdots$ represents the contribution associated with frequency renormalization in this case. Their sum will tell us the net energy flow between the oscillator and the bath, 
\begin{equation}\label{E:ngkje}
	P_{\xi}(\tau)+P_{\gamma}(\tau)=\frac{e}{2}\,\langle\bigl\{\hat{\phi}(\mathbf{z},\tau),\,\dot{\hat{\chi}}(\tau)\bigr\}\rangle+\cdots\,.
\end{equation}
In~\cite{NLFDR}, we show, by the functional method  {without resort to any perturbative argument}, that the righthand side of \eqref{E:ngkje} can be written in terms of the Green's functions of the interacting anharmonic oscillator and those of the free scalar field,
\begin{equation}\label{E:fgbsrhers}
	P_{\xi}(\tau)+P_{\gamma}(\tau)=e^{2}\int_{0}^{\tau}\!ds\;\biggl\{\frac{d}{d\tau}G_{R}^{(\chi)}(\tau,s)\,G_{H,\,\beta}^{(\phi)}(s,\tau)+G_{R,\,0}^{(\phi)}(\tau-s)\frac{d}{d\tau}G_{H}^{(\chi)}(s,\tau)\biggr\}+\cdots\,.
\end{equation}
 {We stress that the Green's function of the oscillator appearing in \eqref{E:fgbsrhers} satisfies an equation of motion like \eqref{E:djsh}, that is, including the self-interaction and the back-actions from the bath field}

The complexity of nonlinear dynamics and the presence  of quantum chaos make it difficult to identify, if possible,  let alone prove, the existence of a stable equilibrium state. In the next section, we will show that at the perturbative level, the dynamics of a weakly anharmonic oscillator, driven by quantum fluctuations of the bath field, does have a stable equilibrium state, if the anharmonic potential obeys certain properties. For now we {\it assume} that the dynamics of the anharmonic oscillator does approach a stable equilibrium state at late times, that the energy exchange between the oscillator and the field will come to a balance and the system settles into an equilibrium state.  Thus we have, {from \eqref{E:fgbsrhers},}
\begin{equation}\label{E:fkbrtjsd}
	\lim_{\tau\to\infty}P_{\xi}(\tau)+P_{\gamma}(\tau)=0
\end{equation}
after subtracting out the contribution corresponding to frequency renormalization. It is shown in~\cite{NLFDR} that, under an additional assumption that the nonstationary components of the anharmonic oscillator's two-point functions have negligible contributions to the integral in \eqref{E:fgbsrhers}, Eq.~\eqref{E:fkbrtjsd} implies an FDR for the anharmonic oscillator
\begin{equation}\label{E:djgeresfjk}
	\overline{G}_{H}^{(\chi)}(\kappa)=\coth\frac{\beta\kappa}{2}\,\operatorname{Im}\overline{G}_{R}^{(\chi)}\!(\kappa)\,,
\end{equation}
with the help of the FDR of the free scalar field
\begin{equation}\label{E:rhfgjdh}
	\overline{G}_{H,\,\beta}^{(\phi)}(\kappa)=\coth\frac{\beta\kappa}{2}\,\operatorname{Im}\overline{G}_{R,\,0}^{(\phi)}(\kappa)\,.
\end{equation}
This is a nonperturbative {and powerful statement}, based on nonequilibrium dynamics~\cite{AFM0}, and thus its validity goes beyond the confines of the conventional FDR based on linear response theory~\cite{LRT}. It is worth noting that the second assumption in general does not hold even if the two-point functions of the anharmonic oscillator become stationary at late times (in comparison, for a linear oscillator, these aforementioned assumptions can be directly shown to hold~\cite{QTD}). Nonetheless, the existence of the FDR \eqref{E:djgeresfjk} provides useful insight in the framework of quantum open systems into issues between equilibration, energy flow balance, and stationarity in the dynamics of an anharmonic oscillator coupled to a field bath.

In~\cite{NLFDR},  the plausibility of these two assumptions used in the nonperturbative arguments are further discussed, based on the perturbative approach to first order in the nonlinear potential.  {Specifically, there we argue that the nonstationary components of the anharmonic oscillator's Green's functions indeed give  exponentially smaller contributions than the stationary component. It  necessitates one of the assumptions used in the nonperturbative arguments, and implies broader validity of that assumption, as a consequence of the damping and small fluctuating driving force associated with the backreaction of the bath field.} 

\section{First perturbative order calculation of energy exchange}\label{S:ertbfkjd}

In this section we present a detailed calculation of the energy exchange between the anharmonic oscillator and the bath field, to first-order in the perturbation of the anharmonic potential. We will show from the dynamical evolution of the reduced system that this energy exchange will be balanced at late times, thus pointing to the existence of an asymptotic equilibrium state.

\begin{figure}
\centering
    \scalebox{0.6}{\includegraphics{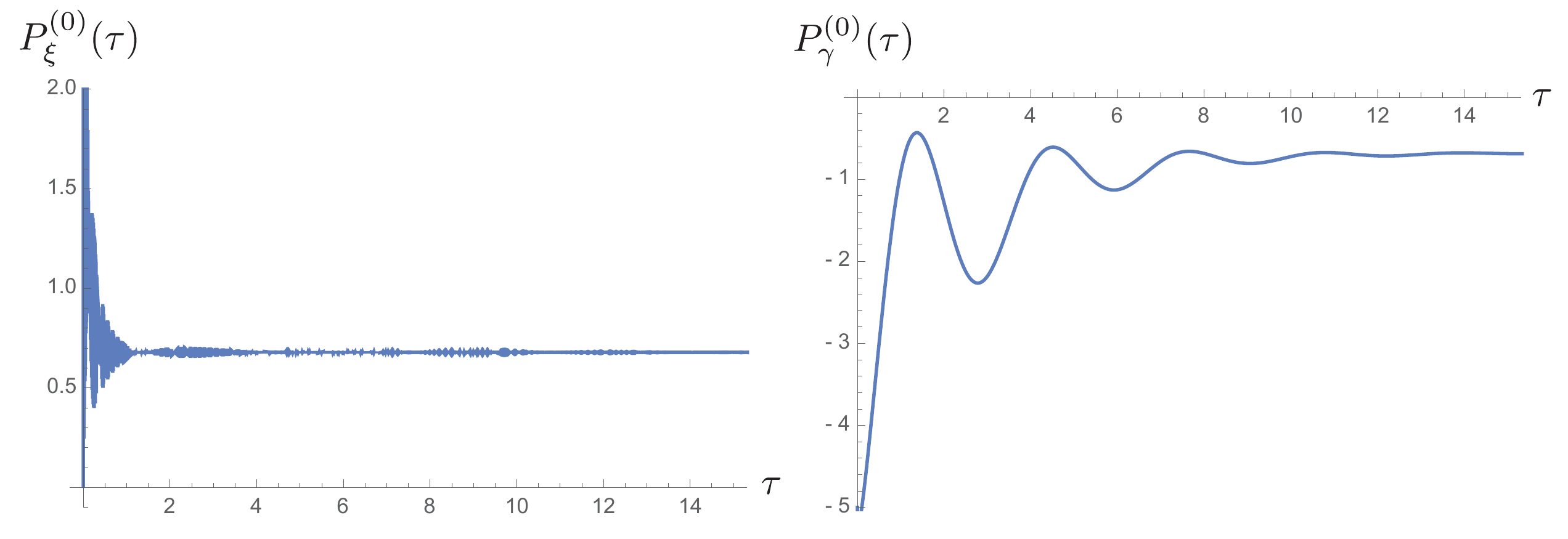}}
    \caption{Time variation of the energy flows in two different channels for the zeroth-order dynamics of the (harmonic) oscillator: $P^{(0)}_{\xi}(\tau)$ is the energy pumped by the fluctuating force in \eqref{E:gbkftes}, and $P^{(0)}_{\gamma}(\tau)$ is the energy dissipated into the bath field. The curves and the relevant parameters are plotted with respect to the appropriate scale such that $\omega=1$, $\gamma=0.2$, $\sigma=0.2$, $\beta=1$, $m=1$, and $\Lambda=500$. The meanings of these parameters can be found in the text.}\label{Fi:PowEx}
\end{figure}

{Since we will take a perturbative approach to discuss the relaxation process by which the energy exchange reaches balance, it would be physically more intuitive if we first briefly outline  this process for the zeroth-order dynamics. It has been discussed~\cite{QTD} in great detail  that the damping due to the self-force, as a consequence of radiation field emitted from the detector, plays an important role. On the one hand, it dissipates the motion of the oscillator by releasing energy back to the environment. On the other hand, it can suppress the nonstationary component of the two-point functions of the oscillator, which in general is not invariant under time translation, a feature of nonequilibrium dynamics. Thus, the two-point functions of the oscillator can possibly become stationary only when both times are greater than the relaxation time. Before that, the nonstationarity of the dynamics is reflected in the time variance of the energy exchange. For example, in Fig.~\ref{Fi:PowEx}, we plot the time variation of the energy exchange for an oscillator initially prepared in a Gaussian wavepacket state with width of the order $\sigma$ and the scalar field initially in its thermal state at temperature $\beta^{-1}$. Except for an initial jolt within a time scale about the inverse of the cutoff frequency $\Lambda^{-1}$, the energy flow $P^{(0)}_{\xi}(\tau)$ supplied by  thermal fluctuations of the quantum field bath  quickly settles down to a positive saturated values. It shows  that the oscillator continually draws energy at a constant rate through this channel. On the other hand, in the same plot, we see that the oscillator also dissipates energy back to the bath field, in response to its motion driven by field fluctuations. The curve of $P^{(0)}_{\gamma}(\tau)$, after a time of the order of the relaxation time $\gamma^{-1}$, gradually approaches a fixed negative value. This value exactly cancels the positive value of $P^{(0)}_{\xi}(\tau)$ at late time. This cancellation is explicitly seen in Fig.~\ref{Fi:PowBal}. We see that the curve  stays at negative values most of the time, meaning that, in this case, before relaxation, the oscillator loses energy faster than it acquires from the bath, but the deficit in energy exchange eases down in the end. With a different set  of initial configurations of the oscillator and the quantum field it can also happen that the oscillator absorbs energy faster than it dissipates. }
\begin{figure}
\centering
    \scalebox{0.8}{\includegraphics{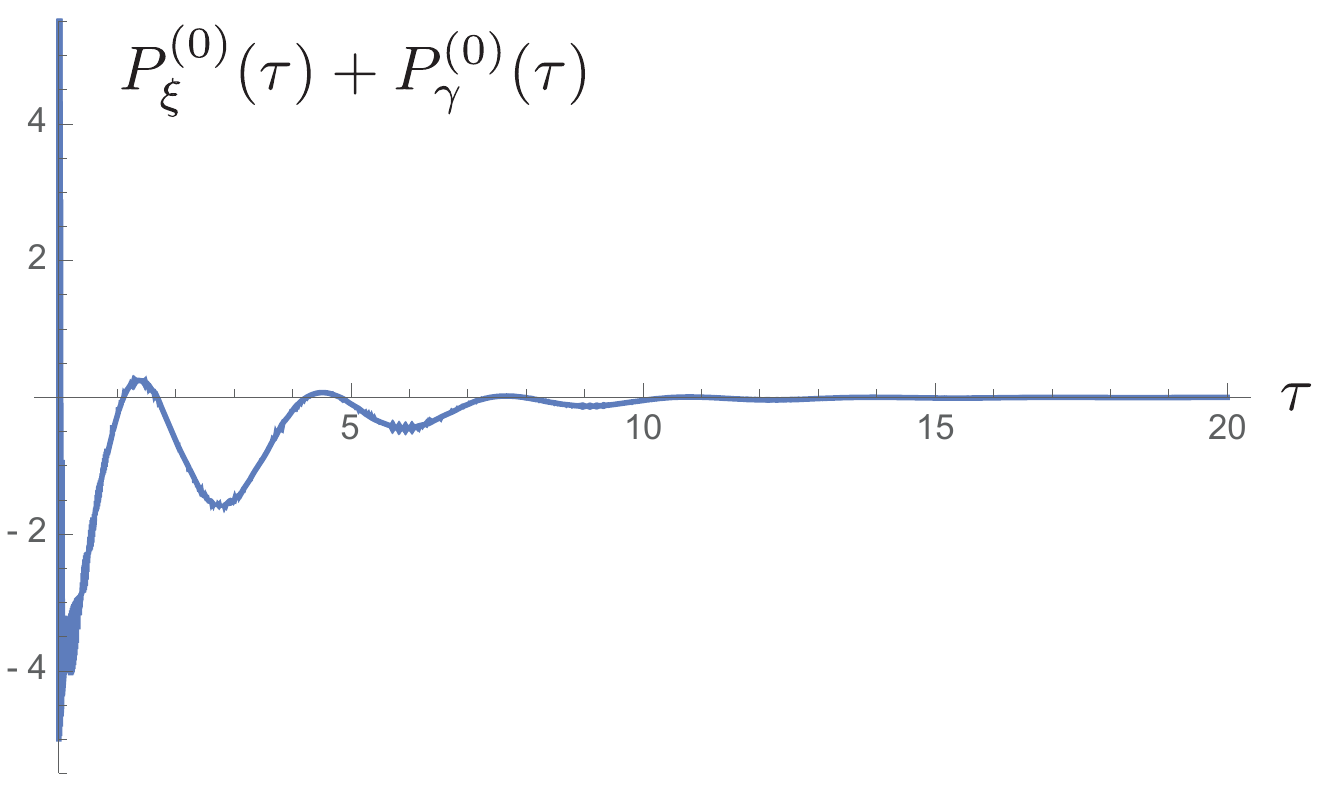}}
    \caption{{Here we show the net energy exchange between the the oscillator and the bath field. It approaches a vanishing value after the evolution time $\tau$ is greater than the relaxation time $\gamma^{-1}$. The parameters take on the same values as in Fig.~\ref{Fi:PowEx}.}}\label{Fi:PowBal}
\end{figure}

{We now turn to a calculation of the first-order corrections of the energy exchange  between the anharmonic oscillator and the scalar field. They are given by
\begin{align}
	P^{(1)}_{\xi}(\tau)&=e^{2}\int_{0}^{t}\!ds\;\Bigl[\frac{d}{d\tau}G_{R,1}^{(\chi)}(\tau,s)\Bigr]\,G_{H,\beta}^{(\phi)}(s,\tau)\,,\\
	P^{(1)}_{\gamma}(\tau)&=e^{2}\int_{0}^{\tau}\!ds\;G_{R,0}^{(\phi)}(\tau-s)\,\Bigl[\frac{d}{d\tau}G_{H,1}^{(\chi)}(s,\tau)\Bigr]\,,
\end{align}
where the first-order corrections of the Green's functions of the anharmonic oscillator are
\begin{align*}
	G_{R,1}^{(\chi)}(\tau,\tau')&=-\frac{\lambda}{2!}\int_{0}^{\tau}\!ds\;G_{R,0}^{(\chi)}(\tau-s)G_{H,0}^{(\chi)}(s,s)G_{R,0}^{(\chi)}(s-\tau')\,,\\
	G_{H,1}^{(\chi)}(\tau,\tau')&=-\frac{\lambda}{2!}\int_{0}^{t}\!ds\;\Bigl[G_{R,0}^{(\chi)}(\tau-s)G_{H,0}^{(\chi)}(s,s)G_{H,0}^{(\chi)}(s,\tau')+G_{R,0}^{(\chi)}(\tau'-s)G_{H,0}^{(\chi)}(s,s)G_{H,0}^{(\chi)}(s,\tau)\Bigr]\,,
\end{align*}
with $0<\tau,\tau'<t$. As a reminder, we write down the zeroth-order Green's functions of the anharmonic oscillator, 
\begin{align*}
	G_{R,0}^{(\chi)}(\tau-\tau')&=\frac{1}{m}\,D_{2}(\tau-\tau')\,,\\
	G_{H,0}^{(\chi)}(\tau,\tau')&=D_{1}(\tau)D_{1}(\tau')\,\langle\hat{\chi}^{2}(0)\rangle+D_{2}(\tau)D_{2}(\tau')\,\frac{\langle\hat{p}^{2}(0)\rangle}{m^{2}}\notag\\
	&\qquad\qquad\qquad\qquad\qquad\qquad\qquad\qquad+\frac{e^{2}}{m^{2}}\int_{0}^{\tau}\!ds\int_{0}^{\tau'}\!ds'\;D_{2}(\tau-s)D_{2}(\tau'-s')G_{H,\beta}^{(\phi)}(s-s')\,,\\
	G_{R,0}^{(\phi)}(\tau)&=-\frac{\theta(\tau)}{2\pi}\,\partial_{\tau}\delta(\tau)\,,\qquad\quad G_{H,\beta}^{(\phi)}(\tau)=\int_{-\infty}^{\infty}\!\frac{d\kappa}{2\pi}\;\widetilde{G}_{H,\beta}^{(\phi)}(\kappa)\,e^{-i\kappa\tau}\,,\qquad\quad \widetilde{G}_{H,\beta}^{(\phi)}(\kappa)=\frac{\kappa}{4\pi}\,\coth\frac{\beta\kappa}{2}\,,\\
	D_{1}(\tau)&=e^{-\gamma\tau}\Bigl(\cos\Omega\tau+\frac{\gamma}{\Omega}\,\sin\Omega\tau\Bigr)\,,\qquad\qquad\quad 
	D_{2}(\tau)=\frac{1}{\Omega}\,e^{-\gamma\tau}\sin\Omega\tau\,,\qquad\qquad\quad\Omega=\sqrt{\omega^{2}-\gamma^{2}}\,.
\end{align*}
Here $\Omega$ gives the resonance frequency. In particular, we explicitly see that the oscillator's Hadamard function is not time translationally invariant, and note that it contains two distinct components: the intrinsic (or active) component depending on the initial condition and decaying with time, and the induced (or passive) component resulting from the fluctuating force and surviving at late time. Next we evaluate the first-order corrections of $P_{\xi}(\tau)$ and $P_{\gamma}(\tau)$ at late times, and examine whether they reach constant values then. If it is the case, then this signifies the existence of an asymptotic equilibrium state, within the validity of perturbation theory.}

{We consider a $\chi^{4}$ anharmonic potential with $\lambda>0$. Since the anharmonic oscillator is driven purely by quantum fluctuations from the bath field, without a large external classical force, the displacement of the oscillator is expected to be quite small. Thus it   cannot effectively be amplified by a weak anharmonic potential. In addition, the frictional force will further damp down the oscillator's motion, so it is not unreasonable to expect that this weakly anharmonic oscillator, when coupled to the thermal bath field, may have well-behaved late-time behavior.}

\subsection{$P^{(1)}_{\xi}(\tau)$}

{First we calculate $P^{(1)}_{\xi}(\tau)$. Since it involves the first order correction of the oscillator's retarded Green's function $G_{R,1}^{(\chi)}(\tau,\tau')$, we observe that the defining integral of $G_{R,1}^{(\chi)}(\tau,\tau')$ contains $G_{H,0}^{(\chi)}(s,s)$. As discussed earlier, it is convenient to decompose it into the intrinsic (active) and the induced (passive) components:
\begin{align}
	\Bigl[G_{H,0}^{(\chi)}(s,s)\Bigr]_{a}&=D_{1}^{2}(s)\,\langle\hat{\chi}^{2}(0)\rangle+D_{2}^{2}(s)\,\frac{\langle\hat{p}^{2}(0)\rangle}{m^{2}}\,,\\
	\Bigl[G_{H,0}^{(\chi)}(s,s)\Bigr]_{p}&=\frac{e^{2}}{m^{2}}\int_{0}^{s}\!da\int_{0}^{s}\!db\;D_{2}(a)D_{2}(b)G_{H,\beta}^{(\phi)}(a-b)\,.
\end{align}
The component $P^{(1)}_{\xi}(\tau)$ associated with the intrinsic (or active, denoted with a subscript $a$) component  is given by 
\begin{align}
	\Bigl[P^{(1)}_{\xi}(\tau)\Bigr]_{a}=e^{2}\int_{0}^{\tau}\!ds\;\Bigl[\frac{d}{d\tau}G_{R,1}^{(\chi)}(\tau,s)\Bigr]_{a}\,G_{H,\beta}^{(\phi)}(s,\tau)\,.
\end{align}
It turns out this component has exponentially small contribution to $P^{(1)}_{\xi}(\tau)$. It decays to zero at least as fast as $\bigl[P^{(1)}_{\xi}(\tau)\bigr]_{a}\sim e^{-\gamma\tau}$ as $\tau\to\infty$.}

{The part of $P^{(1)}_{\xi}(\tau)$ associated with the induced (or passive, denoted with a subscript $p$) component of $G_{H,0}^{(\chi)}(s,s)$ is given by
\begin{align}
	\Bigl[P^{(1)}_{\xi}(\tau)\Bigr]_{p}&=e^{2}\int_{0}^{\tau}\!ds\;\Bigl[\frac{d}{d\tau}G_{R,1}^{(\chi)}(\tau,s)\Bigr]_{p}\,G_{H,\beta}^{(\phi)}(s,\tau)\notag\\
	&=-\frac{\lambda}{2!}\,\frac{e^{4}}{m^{4}}\int_{-\infty}^{\infty}\!\frac{d\kappa}{2\pi}\;\widetilde{G}_{H,\beta}^{(\phi)}\int_{-\infty}^{\infty}\!\frac{d\varpi}{2\pi}\;\widetilde{G}_{H,\beta}^{(\phi)}(\varpi)\int_{0}^{\tau}\!ds\;e^{-i\kappa(s-\tau)}\notag\\
	&\qquad\qquad\qquad\qquad\times\frac{d}{d\tau}\int_{s}^{\tau}\!ds'\;D_{2}(\tau-s')D_{2}(s'-s)\int_{0}^{s'}\!da\int_{0}^{s'}\!db\;D_{2}(a)D_{2}(b)\,e^{-i\varpi(a-b)}\notag\\
	&=-\frac{\lambda}{2!}\,e^{4}\biggl[\int_{-\infty}^{\infty}\!\frac{d\varpi}{2\pi}\;\widetilde{G}_{H,\beta}^{(\phi)}(\varpi)\widetilde{G}_{R,0}^{(\chi)}(\varpi)\widetilde{G}_{R,0}^{(\chi)*}(\varpi)\biggr]\biggl[\int_{-\infty}^{\infty}\!\frac{d\kappa}{2\pi}\;-i\kappa\,\widetilde{G}_{H,\beta}^{(\phi)}(\kappa)\widetilde{G}_{R,0}^{(\chi)2}(\kappa)\biggr]+\cdots\,,\label{E:itrgisd}
\end{align}
where $\cdots$ represents the contributions that decay at least as fast as $e^{-\gamma\tau}$ as $\tau\gg\gamma^{-1}$. We note that both integrals in \eqref{E:itrgisd} will have time-independent values, so it implies that $P^{(1)}_{\xi}(\tau)$ does reach a fixed value at late times. In this calculation, we do not need to make any assumption about the contributions from the nonstationary components of the oscillator's two-point functions, as we do need for the nonperturbative arguments in~\cite{NLFDR}. As is clear from a more explicit analysis in~\cite{NLFDR}, this results from the fact that their contributions are suppressed to be exponentially small at late times by the damping mechanism.} 

\begin{figure}
\centering
    \scalebox{0.55}{\includegraphics{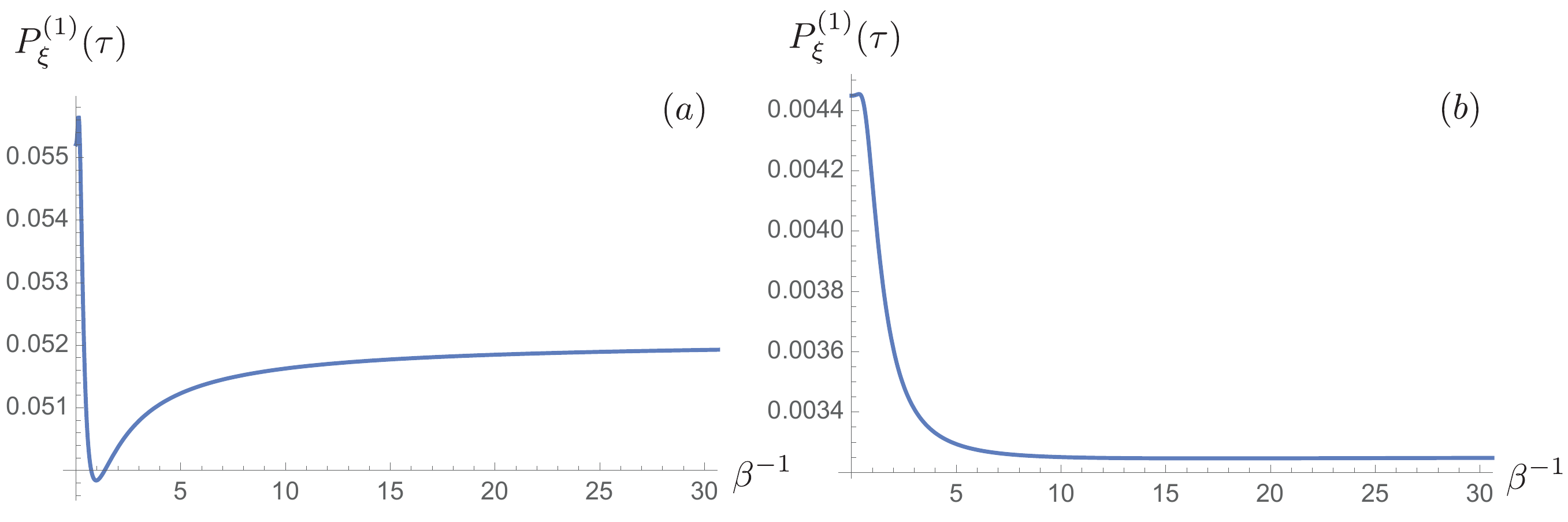}}
    \caption{These plots show  the dependence of $P^{(1)}_{\xi}(\infty)$ at late times on the bath temperature $\beta^{-1}$.  The plots are rescaled, and the parameters are the same as in Fig.~\ref{Fi:PowEx}, except in (a) the renormalized oscillating frequency $\omega=1$ and in (b) $\omega=4$.}\label{Fi:PowHT}
\end{figure}
{For an arbitrary initial temperature of the bath, there is no closed analytical result for either integral in \eqref{E:itrgisd}. We present  the numerical evaluation of $P^{(1)}_{\xi}(\tau)$ in Fig.~\ref{Fi:PowHT}. Thus we first choose a special case that the bath scalar field is initially in its vacuum state, that is, a zero-temperature bath, $\beta\to\infty$, and carry out the corresponding integrals in \eqref{E:itrgisd}. 

\paragraph{Zero-temperature field bath} Note that the zero temperature limit of $\widetilde{G}_{H,\beta}^{(\phi)}(\kappa)$ is
\begin{align}
	\widetilde{G}_{H,\beta}^{(\phi)}(\kappa)&=\frac{\kappa}{4\pi}\,\operatorname{sgn}(\kappa)\,,&&\text{instead of} &\widetilde{G}_{H,\beta}^{(\phi)}(\kappa)&=\frac{\kappa}{4\pi}\,.
\end{align}
The integrals that constitute the nonvanishing contributions in \eqref{E:itrgisd} are respectively given by
\begin{align}
	\int_{-\infty}^{\infty}\!\frac{d\varpi}{2\pi}\;\widetilde{G}_{H,\beta}^{(\phi)}(\varpi)\widetilde{D}_{2}(\varpi)\widetilde{D}_{2}^{*}(\varpi)&=\frac{1}{32\pi\gamma\Omega}\biggl[1+\frac{2}{\pi}\tan^{-1}\Bigl(\frac{\Omega}{2\gamma}-\frac{\gamma}{2\Omega}\Bigr)\biggr]=\frac{1}{16\pi\gamma\Omega}\biggl[1-\frac{2}{\pi}\tan^{-1}\frac{\gamma}{\Omega}\biggr]\,,\label{E:kdbgkes1}\\
	\int_{-\infty}^{\infty}\!\frac{d\kappa}{2\pi}\;-i\kappa\,\widetilde{G}_{H,\beta}^{(\phi)}(\kappa)\widetilde{D}_{2}^{2}(\kappa)&=\frac{1}{16\pi\Omega^{3}}\biggl\{\frac{2\gamma\Omega}{\pi}-\bigl(\Omega^{2}+\gamma^{2}\bigr)\biggl[1-\frac{2}{\pi}\tan^{-1}\frac{\gamma}{\Omega}\biggr]\biggr\}\,,\label{E:kdbgkes2}
\end{align}
These integrals are exact, so they override  the weak oscillator-bath coupling limitation. Combining \eqref{E:kdbgkes1} and \eqref{E:kdbgkes2}, we find that the first-order correction at late times $P^{(1)}_{\xi}(\infty)$ is given by
\begin{align}
	P^{(1)}_{\xi}(\infty)&=-\frac{\lambda}{2!}\,e^{4}\biggl[\int_{-\infty}^{\infty}\!\frac{d\varpi}{2\pi}\;\widetilde{G}_{H,\beta}^{(\phi)}(\varpi)\widetilde{G}_{R,0}^{(\chi)}(\varpi)\widetilde{G}_{R,0}^{(\chi)*}(\varpi)\biggr]\biggl[\int_{-\infty}^{\infty}\!\frac{d\kappa}{2\pi}\;-i\kappa\,\widetilde{G}_{H,\beta}^{(\phi)}(\kappa)\widetilde{G}_{R,0}^{(\chi)2}(\kappa)\biggr]\label{E:fgrurbfs}\\
		&=-\frac{\lambda}{8m^{2}}\frac{\gamma^{2}}{\Omega^{3}}\biggl[1-\frac{2}{\pi}\tan^{-1}\frac{\gamma}{\Omega}\biggr]\biggl[\frac{2}{\pi}-\frac{\Omega^{2}+\gamma^{2}}{\gamma\Omega}\,\biggl(1-\frac{2}{\pi}\tan^{-1}\frac{\gamma}{\Omega}\biggr)\biggr]\,,\label{E:flgruerd}
\end{align}
when $\tau\gg\gamma^{-1}$. 
We note that \eqref{E:flgruerd} gives a positive value and the self-coupling strength $\lambda$ has a dimension $[\lambda]=L^{-5}$.}

\paragraph{High temperature field bath}

{We next check the high-temperature case. We will use the result in \eqref{E:itrgisd}
\begin{align}
	\Bigl[P^{(1)}_{\xi}(\tau)\Bigr]_{p}&=-\frac{\lambda}{2!}\,e^{4}\biggl[\int_{-\infty}^{\infty}\!\frac{d\varpi}{2\pi}\;\widetilde{G}_{H,\beta}^{(\phi)}(\varpi)\widetilde{G}_{R,0}^{(\chi)}(\varpi)\widetilde{G}_{R,0}^{(\chi)*}(\varpi)\biggr]\biggl[\int_{-\infty}^{\infty}\!\frac{d\kappa}{2\pi}\;-i\kappa\,\widetilde{G}_{H,\beta}^{(\phi)}(\kappa)\widetilde{G}_{R,0}^{(\chi)2}(\kappa)\biggr]+\cdots\,.
\end{align}
The evaluation of the first integral
\begin{equation*}
	I_{1}=\int_{-\infty}^{\infty}\!\frac{d\varpi}{2\pi}\;\widetilde{G}_{H,\beta}^{(\phi)}(\varpi)\widetilde{G}_{R,0}^{(\chi)}(\varpi)\widetilde{G}_{R,0}^{(\chi)*}(\varpi)\,,
\end{equation*}
is pretty straightforward. The high temperature limit of $\coth\beta\varpi/2$ is 
\begin{equation}\label{E:fbkbgdh}
	\widetilde{G}_{H,\beta}^{(\phi)}(\varpi)\sim\frac{1}{2\pi\beta}+\frac{\beta\kappa^{2}}{24\pi}+\mathcal{O}(\beta^{3})\,,
\end{equation}
so the leading contribution of the corresponding integral gives
\begin{align}\label{E:fhrt6ytdbd}
	I_{1}=\frac{1}{2\pi\beta}\int_{-\infty}^{\infty}\!\frac{d\varpi}{2\pi}\;\widetilde{G}_{R,0}^{(\chi)}(\varpi)\widetilde{G}_{R,0}^{(\chi)*}(\varpi)=\frac{1}{8\pi m^{2}\beta\gamma}\frac{1}{\Omega^{2}+\gamma^{2}}\,,
\end{align}
for $\beta\Omega<2$. This is consistent with the observation that $I_{1}$ is, apart from the proportionality constant, essentially $\langle\hat{\chi}^{2}(\tau)\rangle^{(0)}$, which at high temperature will be proportional to the bath temperature $\beta^{-1}$ according to the equi-partition theorem.}

The evaluation of the other integral is trickier. Na\"ively, if we use the same approximation scheme \eqref{E:fbkbgdh}, we find that
\begin{align}
	I_{2}=\int_{-\infty}^{\infty}\!\frac{d\kappa}{2\pi}\;-i\kappa\,\widetilde{G}_{H,\beta}^{(\phi)}(\kappa)\widetilde{G}_{R,0}^{(\chi)2}(\kappa)&=\frac{1}{2\pi m^{2}\beta}\int_{-\infty}^{\infty}\!\frac{d\kappa}{2\pi}\;\frac{4\gamma\kappa^{2}(-\kappa^{2}+\omega^{2})}{[(\kappa^{2}-\omega^{2})^{2}+4\gamma^{2}\kappa^{2}]^{2}}=0\,,
\end{align}
for sufficiently high temperature. However, the numerical evidence in Fig~\ref{Fi:PowHT} shows that it is indeed proportional to $\beta$, a finite value, such that the product of these two integrals is constant in $\beta$ for small $\beta$. It implies that we need to include the higher-order terms in the small $\beta$ expansion, that is, the contribution from $\frac{\beta\kappa^{2}}{24\pi}$ in \eqref{E:fbkbgdh}. We find
\begin{align}\label{E:tyudbx}
	I'_{2}=\frac{\beta}{24\pi}\int_{-\infty}^{\infty}\!\frac{d\kappa}{2\pi}\;\bigl(-i\kappa\bigr)\kappa^{2}\,\widetilde{G}_{R,0}^{(\chi)2}(\kappa)&=-\frac{\beta}{48\pi m^{2}}\,.
\end{align}
It is interesting to note that a more careful analysis shows that
\begin{equation}
	\frac{\beta}{24\pi}\int_{-2/\beta}^{2/\beta}\!\frac{d\kappa}{2\pi}\;\bigl(-i\kappa\bigr)\kappa^{2}\,\widetilde{G}_{R,0}^{(\chi)2}(\kappa)\simeq-\frac{\beta}{48\pi m^{2}}+\mathcal{O}(\beta^{3})
\end{equation}
for $\beta\omega\ll1$. This implies that the major contribution of $I_{2}$ comes from the interval $(-2/\beta,+2/\beta)$, and thus in the high-temperature limit $\beta\to0$, the integral $I'_{2}$ should be sufficiently accurate to represent $I_{2}$, and is consistent with the numerical calculations. Now put \eqref{E:fhrt6ytdbd} and \eqref{E:tyudbx} back to \eqref{E:fgrurbfs}, and we arrive at
\begin{align}
	P^{(1)}_{\xi}(\infty)&\simeq\frac{\lambda}{12m^{2}}\frac{\gamma}{\Omega^{2}+\gamma^{2}}\,.
\end{align}
This result for the first-order correction of the power delivered by the fluctuating force from the bath field is independent of its initial temperature, in contrast to the zeroth-order case where it grows linearly with the temperature in the same high-temperature limit. Thus, it appears that anharmonicity {is unable to} amplify the motion of the oscillator driven by the high-temperature thermal noise force from the bath field. 
It imply that at least to the first order perturbation theory, this result may be valid for a wide range of high temperatures, so the bath is not restricted to be a low-temperature thermal field.

{Another interesting feature is that the value of $P^{(1)}_{\xi}(\infty)$ at high temperature seems to be always smaller than its zero temperature value. We see roughly the ratio between the value at the zero temperature and at high temperatures is $3/2$ for small $\gamma/\Omega$ ratio, that is, the weak oscillator-bath coupling limit. Maybe this difference is a partial reflection of the loss of coherence due to the thermal fluctuations of the bath.}

\subsection{$P^{(1)}_{\gamma}(\tau)$}

{Next we compute $P^{(1)}_{\gamma}(\tau)$ at late times. We first re-organize its expressions
\begin{align}
	P^{(1)}_{\gamma}(\tau)&=-\frac{\lambda}{2!}\frac{e^{2}}{2\pi}\int_{0}^{\tau}\!ds\;\partial_{s}\delta(\tau-s)\,\frac{d}{d\tau}G_{H,1}^{(\chi)}(s,\tau)\notag\\
	&=+\frac{\lambda}{2!}\frac{e^{2}}{2\pi}\int_{0}^{\tau}\!ds\;\Bigl[\frac{d}{d\tau}G_{R,0}^{(\chi)}(\tau-s)\Bigr]G_{H,0}^{(\chi)}(s,s)\Bigl[\frac{d}{d\tau}G_{H,0}^{(\chi)}(s,\tau)\Bigr]\,.\label{E:rfbdjhfs}
\end{align}
Here we note that for large $\tau\gg\gamma^{-1}$, the active component of $G_{H,0}^{(\chi)}(s,\tau)$ is exponentially small, so its contribution at late times can be ignored. Thus we focus on the contribution from the passive component $[G_{H,0}^{(\chi)}(s,\tau)]_{p}$. Since there is no obvious simplification for $G_{H,0}^{(\chi)}(s,s)$ in \eqref{E:rfbdjhfs} we need to carry out the calculations to determine the contributions from its active and passive components. The calculations are straightforward, and we find
\begin{align}
	\frac{e^{2}}{2\pi}\int_{0}^{\tau}\!ds\;\Bigl[\frac{d}{d\tau}G_{R,0}^{(\chi)}(\tau-s)\Bigr]\Bigl[D_{1}^{2}(s)\,\langle\hat{\chi}^{2}(0)\rangle+D_{2}^{2}(s)\,\frac{\langle\hat{p}^{2}(0)\rangle}{m^{2}}\Bigr]\Bigl[\frac{d}{d\tau}G_{H,0}^{(\chi)}(s,\tau)\Bigr]_{p}&=\mathcal{O}(e^{-\gamma\tau})\,.
\end{align}
That is, it gives a vanishing contribution at late times.}

{The contribution from the passive component of $G_{H,0}^{(\chi)}(s,s)$ is given by
\begin{align}
	&\quad\frac{e^{2}}{2\pi m}\frac{e^{4}}{m^{4}}\int_{-\infty}^{\infty}\!\frac{d\kappa}{2\pi}\;\widetilde{G}_{H,\beta}^{(\phi)}(\kappa)\int_{-\infty}^{\infty}\!\frac{d\varpi}{2\pi}\;\widetilde{G}_{H,\beta}^{(\phi)}(\varpi)\int_{0}^{\tau}\!ds\;\dot{D}_{2}(\tau-s)\int_{0}^{s}\!da\int_{0}^{s}\!db\;D_{2}(a)\dot{D}_{2}(b)\,e^{-i\varpi(a-b)}\notag\\
	&\qquad\qquad\qquad\qquad\qquad\qquad\qquad\qquad\qquad\qquad\qquad\qquad\qquad\times\int_{0}^{s}\!da\int_{0}^{\tau}\!db\;D_{2}(a)\dot{D}_{2}(b)\,e^{+i\kappa(\tau-s+a-b)}\notag\\
	&=\frac{e^{2}}{2\pi}\,e^{4}\biggl[\int_{-\infty}^{\infty}\!\frac{d\kappa}{2\pi}\;\kappa^{2}\widetilde{G}_{H,\beta}^{(\phi)}(\kappa)\widetilde{G}_{R,0}^{(\chi)2}(\kappa)\widetilde{G}_{R,0}^{(\chi)*}(\kappa)\biggr]\biggl[\int_{-\infty}^{\infty}\!\frac{d\varpi}{2\pi}\;\kappa^{2}\widetilde{G}_{H,\beta}^{(\phi)}(\varpi)\widetilde{G}_{R,0}^{(\chi)}(\varpi)\widetilde{G}_{R,0}^{(\chi)*}(\varpi)\biggr]+\cdots\,.\label{E:djshebjhewe}
\end{align}
It has a structure similar to \eqref{E:fgrurbfs}, so we attempt to re-write the $\kappa$-integral in \eqref{E:djshebjhewe}. We first note that by the FDRs of the free oscillator and the free field, we have
\begin{equation}
	\operatorname{Im}\widetilde{G}_{R,0}^{(\chi)}(\kappa)=e^{2}\Bigl[\operatorname{Im}\widetilde{G}_{R,0}^{(\phi)}(\kappa)\Bigr]\widetilde{G}_{R,0}^{(\chi)}(\kappa)\widetilde{G}_{R,0}^{(\chi)*}(\kappa)\,.
\end{equation}
Thus the $\kappa$-integral in \eqref{E:djshebjhewe} becomes
\begin{align}
	\int_{-\infty}^{\infty}\!\frac{d\kappa}{2\pi}\;\kappa^{2}\,\widetilde{G}_{H,\beta}^{(\phi)}(\kappa)\widetilde{G}_{R,0}^{(\chi)2}(\kappa)\widetilde{G}_{R,0}^{(\chi)*}(\kappa)&=e^{-2}\int_{-\infty}^{\infty}\!\frac{d\kappa}{2\pi}\;\kappa^{2}\,\coth\frac{\beta\kappa}{2}\Bigl[\operatorname{Im}\widetilde{G}_{R,0}^{(\chi)}(\kappa)\Bigr]\Bigl[\operatorname{Re}\widetilde{G}_{R,0}^{(\chi)}(\kappa)\Bigr]\,,
\end{align}
because $\operatorname{Re}\widetilde{G}_{R,0}^{(\chi)}(\kappa)$ and $\operatorname{Im}\widetilde{G}_{R,0}^{(\chi)}(\kappa)$ are respectively even and odd functions of $\kappa$. We arrive at 
\begin{align}
	P^{(1)}_{\gamma}(\infty)&=\frac{\lambda}{2!}\frac{e^{4}}{2\pi}\biggl\{\int_{-\infty}^{\infty}\!\frac{d\kappa}{2\pi}\;\kappa^{2}\,\coth\frac{\beta\kappa}{2}\Bigl[\operatorname{Im}\widetilde{G}_{R,0}^{(\chi)}(\kappa)\Bigr]\Bigl[\operatorname{Re}\widetilde{G}_{R,0}^{(\chi)}(\kappa)\Bigr]\biggr\}\notag\\
	&\qquad\qquad\qquad\qquad\qquad\qquad\qquad\qquad\times\biggl\{\int_{-\infty}^{\infty}\!\frac{d\varpi}{2\pi}\;\kappa^{2}\widetilde{G}_{H,\beta}^{(\phi)}(\varpi)\widetilde{G}_{R,0}^{(\chi)}(\varpi)\widetilde{G}_{R,0}^{(\chi)*}(\varpi)\biggr\}\,.\label{E:gkbsrtgd}
\end{align}
Comparing \eqref{E:gkbsrtgd} with $P^{(1)}_{\xi}(\infty)$ in \eqref{E:fgrurbfs}, we re-write the $\kappa$-integral in \eqref{E:fgrurbfs} into
\begin{align}
	\int_{-\infty}^{\infty}\!\frac{d\kappa}{2\pi}\;-i\kappa\,\widetilde{G}_{H,\beta}^{(\phi)}(\kappa)\widetilde{G}_{R,0}^{(\chi)2}(\kappa)&=\frac{1}{2\pi}\int_{-\infty}^{\infty}\!\frac{d\kappa}{2\pi}\;\frac{\kappa^{2}}{4\pi}\,\coth\frac{\beta\kappa}{2}\,\Bigl[\operatorname{Im}\widetilde{G}_{R,0}^{(\chi)}(\kappa)\Bigr]\Bigl[\operatorname{Re}\widetilde{G}_{R,0}^{(\chi)}(\kappa)\Bigr]\,,
\end{align}
because the product $\bigl[\operatorname{Im}\widetilde{G}_{R,0}^{(\chi)}(\kappa)\bigr]\bigl[\operatorname{Re}\widetilde{G}_{R,0}^{(\chi)}(\kappa)\bigr]$ is an odd function of $\kappa$. After a brute-force calculation, we find
\begin{equation}\label{E:gkdbksdfs}
	P^{(1)}_{\gamma}(\infty)=-P^{(1)}_{\xi}(\infty)\,,
\end{equation}
at late times. There is no need to repeat the evaluation of $P^{(1)}_{\gamma}(\infty)$, because it is given by the negative of \eqref{E:flgruerd}. It is interesting to note that the $\varpi$-integral in $P^{(1)}_{\xi}(\infty)$ and $P^{(1)}_{\gamma}(\infty)$
\begin{equation}
	\int_{-\infty}^{\infty}\!\frac{d\varpi}{2\pi}\;\widetilde{G}_{H,\beta}^{(\phi)}(\varpi)\widetilde{G}_{R,0}^{(\chi)}(\varpi)\widetilde{G}_{R,0}^{(\chi)*}(\varpi)
\end{equation}
is proportional to $\langle\chi^{2}(\infty)\rangle$. Indeed \eqref{E:gkdbksdfs} tells us that from the viewpoint of the anharmonic oscillator, the exchange rate of the energy with the bath field is asymptotically balanced. From the argument sketched in the previous section, this balance of the energy flow will also lead to an FDR for the anharmonic oscillator. This, on the one hand, endows the FDR with a dynamical interpretation, thanks to a full nonequilibrium treatment, and on the other hand, suggests that the FDR is a categorical relation which signifies stable dynamical equilibration.}

\section{summary}

In this paper, we study the nonequilibrium dynamics of a nonlinear quantum open system by the functional method. Specifically we are interested in the dynamical evolution of a weakly anharmonic oscillator linearly coupled to a quantum field, initially in a thermal state. {We seek the evidence that such a dynamical system possesses a stable equilibrium state at late times and describe the scenario for its equilibration.} The back-action  of the fluctuating force from the bath field generates a frictional force in the motion of the system, the weakly anharmonic oscillator. Due to the presence of the nonlinear potential, the reduced open system is non-Gaussian and its dynamics cannot be solved exactly analytically. Thus we resort to a perturbative treatment, where the advantages of the functional method become helpful.

We first construct the `in-in' generating functional for a linear quantum open system, with the system of interest driven by external sources. The variation of this generating functional with respect to the external sources will enable us to derive the physical quantities of the nonlinear system perturbatively. In the examples we presented  the advantages of the functional method begin to show up in the more complicated higher-loop  calculations. Generically speaking, it allows us to write the physical quantities of interest  in terms of the zero-order Green's functions, forming a Wick-like expansion for any initial Gaussian state of the oscillator. {Since real time dynamics needs an `in-in' formulation, where four types of Green's functions are involved instead of just one, the  Feynman propagator, in the `in-out' formalism of, say scattering theory, the advantages of the functional method are amplified.}

Of particular interest is the time-evolution of the first-order corrections of the retarded Green's function (aka dissipation kernel) and the Hadamard function (noise kernel) of the anharmonic oscillator. During the transient period, the nonequilibrium nature of the coupled systems indicates that these two Green's functions are not invariant in time translation. However, for the case of weak anharmonicity, the backactions of the bath field  suppresses the nonstationary component, enabling the first-order (anharmonic) corrections and the zeroth-order (harmonic) contributions of these two Green's functions to be asymptotically stationary. With this, we can identify a FDR between the first-order corrections of the retarded Green's function and the Hadamard function.  Interestingly this FDR has the same form as the linear harmonic oscillator, coupled to a bath field. It is worth mentioning that in the derivation, the coupling between the oscillator and the bath field need not be assumed weak. The identical functional forms of FDR's for the zeroth-order and the first-order corrections motivated us to look into the possibility of a broader nonperturbative argument for  the derivation of the FDR of the anharmonic oscillator coupled to the bath field, and its implications of dynamical equilibration, which is treated in a follow-up paper~\cite{NLFDR}. 

The connection between the existence of the stable equilibrium state and the emergence of the FDR for the linear open systems has been investigated in~\cite{QTD,HHL18,AFM0} by examining the energy exchange between the system oscillator and its environmental bath field. During the nonequilibrium evolution, there is continuous, imbalanced energy flow between these two parties. The quantum fluctuations of the bath field transport energy into the oscillator, while the energy will also be drained back to the field through the oscillator's dissipative dynamics. For the linear open systems, these two energy flow will come to a balance when the evolution reaches a stable equilibrium state.  We thus follow the same line of thought, and calculate the first-order corrections of the energy flows between the anharmonic oscillator and the bath field. We obtain  analytical expressions for these corrections in the special cases when the bath field is at zero temperature and in the high-temperature limit, and present numerical evaluations for the finite-temperature cases. 

Moreover, with the help of the results in~\cite{NLFDR}, we can show that the corrections, due to the nonlinear potential, of the energy flows between the anharmonic oscillator and the bath does have equal values but with opposite sign after the oscillator's dynamics is fully relaxed at a time scale equal to the zeroth-order dynamics; thus implying the existence of the stable equilibrium state within the validity of perturbation theory. This suggests an interesting connection between the dynamical equilibration and the FDR in general. A more general discussion is given in~\cite{NLFDR}. It is also interesting to note that the first-order nonlinear corrections of these energy flows in the high-temperature limit turn out to be independent of the bath temperature. {This may suggest that the high-temperature thermal fluctuations of the bath field will not resonantly drive the nonlinear oscillator, so that there is no need to restrict the bath to be a low temperature field.}  \\

\noindent {\bf Acknowledgments} The bulk of the work presented here (save those in Sec.~III.C and IV) was done when both authors visited Fudan University in 2013 and 2014. We thank Professor Y. S. Wu, Director of the Center for Field Theory and Particle Physics, and Professor J. Shen, Chairman of the physics department, for their kind hospitality. 


\appendix
\section{Derivation of the Classical Values of the Coarse-Grained Effective Action $S_{CG}$}\label{S:btiete}

To carry out the path integral~\cite{GSI88} in \eqref{E:kdjwiqa}, we will need the classical trajectories of $r(s)$ and $q(s)$, which can be obtained by taking the variations of the coarse-grained effective action with respect to $q(s)$ and $r(s)$,
\begin{align}
	 m\,\ddot{r}(s)+m\,\omega_{0}^{2}r(s)-e^{2}\int_{0}^{s}\!ds'\;G_{R}^{(\phi)}(s,s')\,r(s')&=i\,e^{2}\int_{0}^{t}\!ds'\;G_{H,\,\beta}^{(\phi)}(s,s')\,q(s')+j_{r}(s)\,,\label{E:humak}\\
	m\,\ddot{q}(s)+m\,\omega_{0}^{2}q(s)-e^{2}\int_{s}^{t}\!ds'\;G_{R}^{(\phi)}(s',s)\,q(s')&=j_{q}(s)\,.\label{E:jejwja}
\end{align}
They will be used to find the classical value $\overline{S}_{CG}$ of the coarse-grained effective action \eqref{E:lkqoakj},
\begin{align}
	 \overline{S}_{CG}[r,q]&=m\,q(s)\dot{r}(s)\,\Big|_{s=0}^{s=t}-\frac{i}{2}\,e^{2}\int_{0}^{t}\!ds\int_{0}^{t}\!ds'\;q(s)\,G_{H,\,\beta}^{(\phi)}(s,s')\,q(s')+\int_{0}^{t}\!ds\;r(s)j_{q}(s)\,.\label{E:ewuishd}
\end{align}
where $r(s)$, $q(s)$ satisfy the equations of motion. From \eqref{E:humak}, we see that in general $r(s)$ is a complex function with real and imaginary parts, $r(s)=r_{R}(s)+i\,r_{I}(s)$, each of which satisfies
\begin{align}
	 m\,\ddot{r}_{R}(s)+m\,\omega_{0}^{2}r_{R}(s)-e^{2}\int_{0}^{s}\!ds'\;G_{R}^{(\phi)}(s,s')\,r_{R}(s')&=j_{r}(s)\,,\label{E:uiwhka}\\
	 m\,\ddot{r}_{I}(s)+m\,\omega_{0}^{2}r_{I}(s)-e^{2}\int_{0}^{s}\!ds'\;G_{R}^{(\phi)}(s,s')\,r_{I}(s')&=e^{2}\int_{0}^{t}\!ds'\;G_{H,\,\beta}^{(\phi)}(s,s')\,q(s')\,.\label{E:iejwja}
\end{align}
We combine \eqref{E:iejwja} and \eqref{E:jejwja} in such a way
\begin{equation}
	\int_{0}^{t}\!ds\;\Bigl[q(s)\times\text{\eqref{E:iejwja}}-r_{I}(s)\times\text{\eqref{E:jejwja}}\Bigr]\,,
\end{equation}
that it gives
\begin{equation*}
	 m\,q(s)\dot{r}_{I}(s)-m\,r_{I}(s)\dot{q}(s)\,\Big|_{s=0}^{s=t}=e^{2}\int_{0}^{t}\!ds\int_{0}^{t}\!ds'\;q(s)\,G_{H,\,\beta}^{(\phi)}(s,s')\,q(s')-\int_{0}^{t}\!ds\;r_{I}(s)j_{q}(s)\,.
\end{equation*}
Since $r_{b}$ and $r_{a}$, the values of $r(t)$ at the final time $t$ and the initial time $t=0$ respectively, are real, we set $r_{I}(0)=0=r_{I}(t)$ and
\begin{equation}
	 m\,q(s)\dot{r}_{I}(s)\,\Big|_{s=0}^{s=t}=e^{2}\int_{0}^{t}\!ds\int_{0}^{t}\!ds'\;q(s)\,G_{H,\,\beta}^{(\phi)}(s,s')\,q(s')-\int_{0}^{t}\!ds\;r_{I}(s)j_{q}(s)\,.
\end{equation}
Substituting this result into \eqref{E:ewuishd} leads to
\begin{equation}\label{E:wiengha}
	 \overline{S}_{CG}[r,q]=m\,q(s)\dot{r}_{R}(s)\,\Big|_{s=0}^{s=t}+\frac{i}{2}\,e^{2}\int_{0}^{t}\!ds\int_{0}^{t}\!ds'\;q(s)\,G_{H,\,\beta}^{(\phi)}(s,s')\,q(s')+\int_{0}^{t}\!ds\;r_{R}(s)j_{q}(s)\,.
\end{equation}
Next we express \eqref{E:wiengha} in terms of the endpoints or boundary values of $q(s)$ and $r_{R}(s)$. Since the equations of motion for $q(s)$ and $r_{R}(s)$ depend on the sources $j_{q}$ and $j_{r}$, their general solutions will also have contributions from the sources,
\begin{equation}\label{E:iuebjqa}
	r(s)=D_{1}(s)\,r_{a}+D_{2}(s)\,\dot{r}_{a}+\frac{1}{m}\int_{0}^{s}\!ds'\;D_{2}(s-s')\,j_{r}(s')\,,\qquad\qquad t>s>0\,.
\end{equation}
Hereafter, we will suppress the subscript $R$ in $r_{R}(s)$ unless confusion arises.

The fundamental solutions $D_{i}(s)$ are the homogeneous solutions to \eqref{E:uiwhka}, satisfying
\begin{equation*}
	D_{1}(0)=1\,,\qquad\qquad\dot{D}_{1}(0)=0\,,\qquad\qquad D_{2}(0)=0\,,\qquad\qquad\dot{D}_{2}(0)=1\,.
\end{equation*}
As for $q(s)$, it can be related to $r(t-s)$, but since it has a source term in the equation of motion \eqref{E:jejwja}, this correspondence is less transparent. Let us assume that $q(s)$ is given by
\begin{equation}\label{E:haoduw}
	 q(s)=D_{1}(t-s)\,q_{b}+D_{2}(t-s)\,\dot{q}_{b}+\frac{1}{m}\int_{0}^{t-s}\!ds'\;D_{2}(t-s-s')\,j_{q}(t-s')\,,\qquad\qquad t>s>0\,,
\end{equation}
and show that it satisfies \eqref{E:jejwja}. Substituting \eqref{E:haoduw} into the lefthand side of \eqref{E:jejwja}  we have
\begin{align}
	&\quad\; m\,\ddot{q}(s)+m\,\omega_{0}^{2}q(s)-e^{2}\int_{s}^{t}\!ds'\;G_{R}^{(\phi)}(s',s)\,q(s')\notag\\
	&=j_{q}(s)+\int_{0}^{\tau}\!ds''\;\frac{d^{2}}{d\tau^{2}}\,D_{2}(\tau-s'')\,j_{q}(t-s'')+\int_{0}^{\tau}\!ds''\;\omega_{0}^{2}D_{2}(\tau-s'')\,j_{q}(t-s'')\notag\\
	 &\qquad\qquad\qquad\qquad\qquad\qquad\qquad\qquad-\frac{e^{2}}{m}\int_{0}^{t}\!d\tau'\;G_{R}^{(\phi)}(\tau-\tau')\int_{0}^{\tau'}\!ds''\;D_{2}(\tau'-s'')\,j_{q}(t-s'')\,,\label{E:laakqp}
\end{align}
with $\tau=t-s$ and $\tau'=t-s'$. We note that   $t>\tau>\tau'>0$ by construction, and the fundamental solution $D_{2}(s)$ is zero when $s<0$, so the last term in \eqref{E:laakqp} can be written as
\begin{align*}
	 -\frac{e^{2}}{m}\int_{0}^{t}\!d\tau'\;G_{R}^{(\phi)}(\tau-\tau')\int_{0}^{\tau'}\!ds''\;D_{2}(\tau'-s'')\,j_{q}(t-s'')=-\frac{e^{2}}{m}\int_{0}^{\tau}\!d\tau'\;G_{R}^{(\phi)}(\tau-\tau')\int_{0}^{\tau}\!ds''\;D_{2}(\tau'-s'')\,j_{q}(t-s'')\,.
\end{align*}
Therefore we have \eqref{E:laakqp} given by
\begin{align}
	&\quad\; m\,\ddot{q}(s)+m\,\omega_{0}^{2}q(s)-e^{2}\int_{s}^{t}\!ds'\;G_{R}^{(\phi)}(s',s)\,q(s')\label{E:oeuwhnk}\\
	&=j_{q}(s)+\int_{0}^{\tau}\!ds''\left\{\frac{d^{2}}{d\tau^{2}}\,D_{2}(\tau-s'')+\omega_{0}^{2}D_{2}(\tau-s'')-\frac{e^{2}}{m}\int_{0}^{\tau}\!d\tau'\;G_{R}^{(\phi)}(\tau-\tau')D_{2}(\tau'-s'')\right\}j_{q}(t-s'')\,.\notag
\end{align}
It can be shown that the second term in \eqref{E:oeuwhnk} vanishes, and \eqref{E:jejwja} follows. Therefore \eqref{E:haoduw} is indeed the general solution to \eqref{E:jejwja}. For future reference we can also write \eqref{E:haoduw} as
\begin{equation}\label{E:saoduw}
	q(s)=D_{1}(t-s)\,q_{b}+D_{2}(t-s)\,\dot{q}_{b}+\frac{1}{m}\int_{s}^{t}\!ds'\;D_{2}(s'-s)\,j_{q}(s')\,,\qquad t>s,\,s'>0\,.
\end{equation}
Next we express \eqref{E:iuebjqa} and \eqref{E:saoduw} in terms of the endpoints $q_{b}$, $r_{b}$ and $q_{a}$, $r_{a}$, so that it will facilitate calculations of the evolutionary operators used to evaluate the density matrix \eqref{E:ngbrhtdfs}.

After some algebraic manipulation, we have
\begin{align}
	r(s)&=\alpha(s)\,r_{a}+\beta(s)\,r_{b}+J_{r}(s)\,,\label{E:uierda}\\
	q(s)&=\alpha(t-s)\,q_{b}+\beta(t-s)\,q_{a}+J_{q}(s)\,,\label{E:vierda}
\end{align}
where
\begin{align*}
	\alpha(s)&=D_{1}(s)-\frac{D_{1}(t)}{D_{2}(t)}\,D_{2}(s)\,,\qquad\qquad\qquad\qquad\qquad\qquad\beta(s)=\frac{D_{2}(s)}{D_{2}(t)}\,,\\
	 J_{r}(s)&=\frac{1}{m}\int_{0}^{s}\!ds'\;D_{2}(s-s')\,j_{r}(s')-\frac{1}{m}\int_{0}^{t}\!ds'\;\frac{D_{2}(s)}{D_{2}(t)}\,D_{2}(t-s')\,j_{r}(s')\,,\\
	 J_{q}(s)&=\frac{1}{m}\int_{s}^{t}\!ds'\;D_{2}(s'-s)\,j_{q}(s')-\frac{1}{m}\int_{0}^{t}\!ds'\;\frac{D_{2}(s')}{D_{2}(t)}\,D_{2}(t-s)\,j_{q}(s')\,.
\end{align*}
The evolutionary propagator $J(r_{b},q_{b},t;r_{a},q_{a},0)$ is then given by
\begin{equation}
	J(r_{b},q_{b},t;r_{a},q_{a},0)\propto\exp\Big(i\,\overline{S}_{CG}[r,q]\Bigr) \label{J-SCG}
\end{equation}
with
\begin{equation*}
	 \overline{S}_{CG}[r,q]=m\,q(s)\dot{r}_{R}(s)\,\Big|_{s=0}^{s=t}+\frac{i}{2}\,e^{2}\int_{0}^{t}\!ds\int_{0}^{t}\!ds'\;q(s)\,G_{H,\,\beta}^{(\phi)}(s,s')\,q(s')+\int_{0}^{t}\!ds\;r_{}(s)j_{q}(s)\,,
\end{equation*}
with $r(s)$ and $q(s)$ given by \eqref{E:uierda} and \eqref{E:vierda}.

\section{Evaluation of the `In-in' Generating Functional}\label{S:kgbrkt}

It is instructive to give an explicit calculation of the generating functional $\mathcal{Z}[j_{q},j_{r};t)$ when the initial state of the oscillator assumes a Gaussian form. Starting from \eqref{E:kowuqha}, we have
\begin{align*}
	 \mathcal{Z}[j_{q},j_{r};t)&=\frac{m}{2\pi\,D_{2}(t_{b})}\int\!dq_{b}\,dr_{b}\;\delta(q_{b})\int\!dq_{a}\,dr_{a}\;\rho_{\chi}(q_{a},r_{a};0)\,\exp\left\{i\,m\,q(s)\dot{r}(s)\,\Big|_{s=0}^{s=t}\right.\notag\\
	 &\qquad\qquad\qquad\qquad\qquad\qquad\qquad-\left.\frac{e^{2}}{2}\int_{0}^{t}\!ds\int_{0}^{t}\!ds'\;q(s)\,G_{H,\,\beta}^{(\phi)}(s,s')\,q(s')+i\int_{0}^{t}\!ds\;r_{}(s)j_{q}(s)\right\}\,.
\end{align*}
Plugging \eqref{E:uierda} and \eqref{E:vierda} in gives a lengthy expression in terms of the endpoints of $q$ and $r$,
\begin{align}
	 \mathcal{Z}[j_{q},j_{r};t)&=\frac{m}{2\pi\,D_{2}(t_{b})}\int\!dq_{b}\,dr_{b}\int\!dq_{a}\,dr_{a}\;\delta(q_{b})\exp\biggl\{i\,m\,\dot{\alpha}(t)\,q_{b}r_{a}+i\,m\,\dot{\beta}(t)\,q_{b}r_{b}+i\,m\,\dot{J}_{r}(t)\,q_{b}\biggr.\notag\\
	 &\qquad\qquad-i\,m\,\dot{\alpha}(0)\,q_{a}r_{a}-i\,m\,\dot{\beta}(0)\,q_{a}r_{b}-i\,m\,\dot{J}_{r}(0)\,q_{a}+i\,A\,r_{a}+i\,B\,r_{b}+i\,C\label{E:wejowsa}\\
	 &\qquad\qquad-\biggl.I_{11}\,q_{b}^{2}-I_{22}\,q_{a}^{2}-I_{12}\,q_{b}q_{a}-I_{10}\,q_{b}-I_{20}\,q_{a}-I_{00}\biggr\}\times\left(\frac{1}{\pi\sigma^{2}}\right)^{1/2}\exp\biggl\{-\frac{1}{\sigma^{2}}\left[r_{a}^{2}+\frac{1}{4}\,q_{a}^{2}\right]\biggr\}\,,\notag
\end{align}
where we have introduced the following shorthand notations
\begin{align*}
	 \dot{\alpha}(t)&=\dot{D}_{1}(t)-\frac{D_{1}(t)}{D_{2}(t)}\,\dot{D}_{2}(t)\,,\qquad\qquad\dot{\alpha}(0)=-\frac{D_{1}(t)}{D_{2}(t)}\,,\qquad\qquad\dot{\beta}(t)=\frac{\dot{D}_{2}(t)}{D_{2}(t)}\,,\qquad\qquad\dot{\beta}(0)=\frac{1}{D_{2}(t)}\,,\\
	 \dot{J}_{r}(t)&=\frac{1}{m}\int_{0}^{t}\!ds'\left[\dot{D}_{2}(t-s')-\frac{\dot{D}_{2}(t)}{D_{2}(t)}\,D_{2}(t-s')\right]j_{r}(s')\,,\qquad\qquad\dot{J}_{r}(0)=-\frac{1}{m}\int_{0}^{t}\!ds'\;\frac{1}{D_{2}(t)}\,D_{2}(t-s')\,j_{r}(s')\,,\\
	A&=\int_{0}^{t}\!ds'\;\alpha(s')j_{q}(s')\,,\qquad\qquad\qquad B=\int_{0}^{t}\!ds'\;\beta(s')j_{q}(s')\,,\qquad\qquad\qquad C=\int_{0}^{t}\!ds'\;J_{r}(s')j_{q}(s')\,,\\
	I_{11}&=\frac{e^{2}}{2}\int_{0}^{t}\!ds\!\int_{0}^{t}\!ds'\;\alpha(t-s)\,G_{H,\,\beta}^{(\phi)}(s-s')\,\alpha(t-s')\,,\quad I_{22}=\frac{e^{2}}{2}\int_{0}^{t}\!ds\!\int_{0}^{t}\!ds'\;\beta(t-s)\,G_{H,\,\beta}^{(\phi)}(s-s')\,\beta(t-s')\,,\\
	I_{12}&=e^{2}\int_{0}^{t}\!ds\!\int_{0}^{t}\!ds'\;\alpha(t-s)\,G_{H,\,\beta}^{(\phi)}(s-s')\,\beta(t-s')\,,\quad I_{10}=e^{2}\int_{0}^{t}\!ds\!\int_{0}^{t}\!ds'\;\alpha(t-s)\,G_{H,\,\beta}^{(\phi)}(s-s')\,J_{q}(t-s')\,,\\
	I_{20}&=e^{2}\int_{0}^{t}\!ds\!\int_{0}^{t}\!ds'\;\beta(t-s)\,G_{H,\,\beta}^{(\phi)}(s-s')\,J_{q}(t-s')\,,\quad I_{00}=\frac{e^{2}}{2}\int_{0}^{t}\!ds\!\int_{0}^{t}\!ds'\;J_{q}(t-s)\,G_{H,\,\beta}^{(\phi)}(s-s')\,J_{q}(t-s')\,,
\end{align*}
to make the generating functional $\mathcal{Z}[j_{q},j_{r}]$ more manageable. We also assume that the initial density operator of the system is given by
\begin{equation}\label{E:bgriusd}
	 \rho_{\chi}(q_{a},r_{a},0)=\left(\frac{1}{\pi\sigma^{2}}\right)^{1/2}\exp\biggl\{-\frac{1}{\sigma^{2}}\left[r_{a}^{2}+\frac{1}{4}\,q_{a}^{2}\right]\biggr\}\,.
\end{equation}
In fact, it can be any Gaussian form more general than \eqref{E:bgriusd}.

We first perform the integral over $q_{b}$, and then over $r_{b}$, so \eqref{E:wejowsa} becomes
\begin{align}
	 \mathcal{Z}[j_{q},j_{r}]&=\left(\frac{1}{\pi\sigma^{2}}\right)^{1/2}\int\!dq_{a}\,dr_{a}\;\delta\bigl(q_{a}-\frac{B}{m\dot{\beta}(0)}\bigr)\exp\biggl\{-i\,m\,\dot{\alpha}(0)\,q_{a}r_{a}-i\,m\,\dot{J}_{r}(0)\,q_{a}\biggr.\notag\\
	 &\qquad\qquad\qquad\qquad\qquad\qquad\qquad\qquad+\biggl.i\,A\,r_{a}+i\,C-I_{22}\,q_{a}^{2}-I_{20}\,q_{a}-I_{00}-\frac{1}{\sigma^{2}}\left[r_{a}^{2}+\frac{1}{4}\,q_{a}^{2}\right]\biggr\}\,.\notag
\intertext{Carrying out the integral of $q_{a}$ is straightforward and leads to}
	&=\exp\biggl\{-\frac{\sigma^{2}}{4}\Bigl(A-\frac{\dot{\alpha}(0)}{\dot{\beta}(0)}\,B\Bigr)^{2}-\frac{1}{4\sigma^{2}}\Bigl(\frac{B}{m\dot{\beta}(0)}\Bigr)^{2}+i\,\Bigl(C-\frac{\dot{J}_{r}(0)}{\dot{\beta}(0)}\Bigr)\biggr.\\
	 &\qquad\qquad\qquad\qquad\qquad\qquad\qquad\qquad\qquad\qquad\qquad\qquad-\biggl.I_{22}\Bigl(\frac{B}{m\dot{\beta}(0)}\Bigr)^{2}-I_{20}\Bigl(\frac{B}{m\dot{\beta}(0)}\Bigr)-I_{00}\biggr\}\,.\notag
\end{align}
Before proceeding, we first re-write some of the expressions in the exponential,
\begin{align*}
	\frac{B}{m\dot{\beta}(0)}&=\frac{1}{m}\int_{0}^{t}\!ds\;D_{2}(s)\,j_{q}(s)\,,\qquad\qquad A-\frac{\dot{\alpha}(0)}{\dot{\beta}(0)}\,B=\int_{0}^{t}\!ds\;D_{1}(s)\,j_{q}(s)\,,\\
	 C-\frac{\dot{J}_{r}(0)}{\dot{\beta}(0)}\,B&=\frac{1}{m}\int_{0}^{t}\!ds\!\int_{0}^{s}\!ds'\;j_{q}(s)\,D_{2}(s-s')\,j_{r}(s')\,,\\
	I_{22}\Bigl(\frac{B}{m\dot{\beta}(0)}\Bigr)^{2}&+I_{20}\Bigl(\frac{B}{m\dot{\beta}(0)}\Bigr)+I_{00}=\frac{e^{2}}{2}\int_{0}^{t}\!ds\!\int_{0}^{t}\!ds'\;\mathfrak{J}_{q}(s)\,G_{H,\,\beta}^{(\phi)}(s-s')\,\mathfrak{J}_{q}(s')\,,
\end{align*}
where we have defined
\begin{align}
	\mathfrak{J}_{q}(s)&=\beta(t-s)\Bigl(\frac{B}{m\dot{\beta}(0)}\Bigr)+J_{q}(s)\frac{1}{m}\int_{s}^{t}\!ds'\;D_{2}(s'-s)\,j_{q}(s')=\frac{1}{m}\int_{0}^{t}\!ds'\;D_{2}(s'-s)\,j_{q}(s')\,,
\end{align}
due to the fact that $D_{2}(s'-s)=0$ for $s'<s$. Combining these results together gives
\begin{align}\label{E:generating}  
	 \mathcal{Z}[j_{q},j_{r}]&=\exp\biggl\{-\frac{1}{4}\int_{0}^{t}\!ds\!\int_{0}^{t}\!ds'\;j_{q}(s)\Bigl[\sigma^{2}D_{1}(s)D_{1}(s')+\frac{1}{m^{2}\sigma^{2}}\,D_{2}(s)D_{2}(s')\Bigr]j_{q}(s')\biggr.\\
	 &\qquad\qquad+\biggl.\frac{i}{m}\int_{0}^{t}\!ds\!\int_{0}^{s}\!ds'\;j_{q}(s)\,D_{2}(s-s')\,j_{r}(s')-\frac{e^{2}}{2}\int_{0}^{t}\!ds\!\int_{0}^{t}\!ds'\;\mathfrak{J}_{q}(s)\,G_{H,\,\beta}^{(\phi)}(s-s')\,\mathfrak{J}_{q}(s')\biggr\}\,.\notag
\end{align}
The normalization is chosen such that when $j_{q}=0=j_{r}$, we have $\mathcal{Z}[0,0]$=1.

\section{Quantum Statistical Averages in the Linear Model in Sec.~\ref{S:brkfbdrt}}\label{S:eibde}
{In this appendix, we give more examples to show how to use either the density matrix approach or the functional method to find the covariance matrix elements and the two-point functions for the linear oscillator model.}

\subsection{Reduced density matrix approach}
Here we present detailed and explicit calculations of the quantum statistical averages of the covariance matrix elements for the driven, linear oscillator model, based on the reduced density matrix of $\chi$. Evaluating multiple Gaussian integrations is straightforward, even a bit tedious, but they are handy for future references and extensions. More terms are presented in arXiv:1912.12803. Among these calculations, the evaluation of $\langle\,\hat{\chi}_{b}\,\hat{p}_{b}\,\rangle_{j}$ is particularly important because it offers a self-consistent check on the equal-time commutation relations.

\subsubsection{$\langle\,\hat{\chi}_{b}\,\rangle_{j}$}\label{S:enrked}
Here we compute $\langle\,\hat{\chi}_{b}\,\rangle_{j}$ with $j$ collectively representing $\{j_{r},j_{q}\}$. Thus $\langle\,\hat{\chi}_{b}\,\rangle_{j}$ is the quantum expectation value of $\hat{\chi}$ when the external source $j$ is present. The procedure is very similar to the calculations of the generating functional. We start with
\begin{align}
	 \langle\,\hat{\chi}_{b}\,\rangle_{j}\,\mathcal{Z}^{(0)}[j]&=\int\!dq_{b}\,dr_{b}\int\!dq_{a}\,dr_{a}\;\delta(q_{b})\left[r_{b}+\frac{1}{2}\,q_{b}\right]J(q_{b},r_{b},t;q_{a},r_{a},0)\rho_{\chi}(q_{a},r_{a},0)\notag\\
		 &=\frac{m}{2\pi\,D_{2}(t_{b})}\int\!dq_{b}\,dr_{b}\int\!dq_{a}\,dr_{a}\;\delta(q_{b})\left[r_{b}+\frac{1}{2}\,q_{b}\right]\exp\left\{i\,m\,q(s)\dot{r}(s)\,\Big|_{s=0}^{s=t}\right.\notag\\
	 &\qquad\qquad\qquad\qquad\qquad\qquad\quad-\left.\frac{1}{2}\,e^{2}\int_{0}^{t}\!ds\int_{0}^{t}\!ds'\;q(s)\,G_{H,\,\beta}^{(\phi)}(s,s')\,q(s')+i\int_{0}^{t}\!ds\;r_{}(s)j_{q}(s)\right\}\notag\\
	 &\qquad\qquad\qquad\times\left(\frac{1}{\pi\sigma^{2}}\right)^{1/2}\exp\biggl\{-\frac{1}{\sigma^{2}}\left[r_{a}^{2}+\frac{1}{4}\,q_{a}^{2}\right]\biggr\}\notag\\
	 &=\frac{i}{m\dot{\beta}(0)}\left(\frac{1}{\pi\sigma^{2}}\right)^{1/2}\!\int\!dr_{a}\;\Bigl[i\,m\,\dot{\alpha}(0)\,r_{a}+i\,m\,\dot{J}_{r}(0)+2I_{22}\,\bar{q}_{a}+I_{20}+\frac{1}{2\sigma^{2}}\,\bar{q}_{a}\Bigr]\notag\\
	 &\qquad\qquad\qquad\qquad\qquad\qquad\times\exp\biggl\{-\frac{1}{\sigma^{2}}\Bigl[r_{a}-\frac{i}{2}\,\Bigl(A-\frac{\dot{\alpha}(0)}{\dot{\beta}(0)}\,B\Bigr)\,\sigma^{2}\Bigr]^{2}+i\,\Bigl(C-\frac{\dot{J}_{r}(0)}{\dot{\beta}(0)}\Bigr)\biggr.\notag\\
	 &\qquad\qquad\qquad\qquad\qquad\qquad\qquad\qquad\quad-\biggl.\frac{\sigma^{2}}{4}\Bigl(A-\frac{\dot{\alpha}(0)}{\dot{\beta}(0)}\,B\Bigr)^{2}-\frac{1}{4\sigma^{2}}\,\bar{q}_{a}^{2}-I_{22}\bar{q}_{a}^{2}-I_{20}\bar{q}_{a}-I_{00}\biggr\}\,,\notag
\end{align}
where we have used $\bar{q}_{a}$ to denote
\begin{equation*}
	\bar{q}_{a}=\frac{B}{m\dot{\beta}(0)}=\frac{1}{m}\int_{0}^{t}\!ds\;D_{2}(s)\,j_{q}(s)\,.
\end{equation*}
The relevant notations and the initial state of the oscillator have been introduced in Sec.~\ref{S:bgerthsb}. Thus the expectation value of $\hat{\chi}_{b}$ becomes
\begin{align}\label{E:huwyq}
	 \langle\,\hat{\chi}_{b}\,\rangle_{j}\,\mathcal{Z}^{(0)}[j]&=\left\{\frac{1}{m}\int_{0}^{t}\!ds\;D_{2}(t-s)\,j_{r}(s)+\frac{i}{2}\int_{0}^{t}\!ds\;\Bigl[\sigma^{2}D_{1}(t)D_{1}(s)+\frac{1}{m^{2}\sigma^{2}}\,D_{2}(t)D_{2}(s)\Bigr]j_{q}(s)\right.\\
	 &\qquad\qquad\qquad\qquad\qquad\qquad\qquad+\left.i\,\frac{e^{2}}{m}\int_{0}^{t}\!ds\!\int_{0}^{t}\!ds'\;D_{2}(t-s)\,G_{H,\,\beta}^{(\phi)}(s-s')\,\mathfrak{J}_{q}(s')\right\}\,\mathcal{Z}^{(0)}[j]\,,\notag
\end{align}
where we have used
\begin{align}
	 -i\frac{\sigma^{2}}{2}\frac{\dot{\alpha}(0)}{\dot{\beta}(0)}\Bigl(A-\frac{\dot{\alpha}(0)}{\dot{\beta}(0)}\,B\Bigr)&=\frac{i}{2}\int_{0}^{t}\!ds\;\sigma^{2}D_{1}(t)D_{1}(s)\,j_{q}(s)\,,\\
	 \frac{i}{m\dot{\beta}(0)}\times\frac{1}{2\sigma^{2}}\,\bar{q}_{a}&=\frac{i}{2}\int_{0}^{t}\!ds\;\frac{1}{m^{2}\sigma^{2}}\,D_{2}(t)D_{2}(s)\,j_{q}(s)\,,\\
	-\frac{\dot{J}_{r}(0)}{\dot{\beta}(0)}&=\frac{1}{m}\int_{0}^{t}\!ds\;D_{2}(t-s)\,j_{r}(s)\,,\\
	 \frac{i}{m\dot{\beta}(0)}\Bigl[2I_{22}\,\bar{q}_{a}+I_{20}\Bigr]&=i\,\frac{e^{2}}{m}\int_{0}^{t}\!ds\!\int_{0}^{t}\!ds'\;D_{2}(t-s)\,G_{H,\,\beta}^{(\phi)}(s-s')\,\mathfrak{J}_{q}(s')\,.
\end{align}
Therefore we arrive at
\begin{align}
	\langle\,\hat{\chi}_{b}\,\rangle_{j}&=\frac{1}{m}\int_{0}^{t}\!ds\;D_{2}(t-s)\,j_{r}(s)+\frac{i}{2}\int_{0}^{t}\!ds\;\Bigl[\sigma^{2}D_{1}(t)D_{1}(s)+\frac{1}{m^{2}\sigma^{2}}\,D_{2}(t)D_{2}(s)\Bigr]j_{q}(s)\\
	 &\qquad\qquad\qquad\qquad\qquad\qquad\qquad\qquad\qquad\qquad+i\,\frac{e^{2}}{m}\int_{0}^{t}\!ds\!\int_{0}^{t}\!ds'\;D_{2}(t-s)\,G_{H,\,\beta}^{(\phi)}(s-s')\,\mathfrak{J}_{q}(s')\,.\notag
\end{align}
In the case that the external sources $j_{q}$, $j_{r}$ vanish, we recover $\langle\,\hat{\chi}_{b}\,\rangle=0$ for the case of a free harmonic oscillator bilinearly coupled to a massless scalar field initially in a thermal state.

\subsubsection{$\langle\,\hat{\chi}_{b}^{2}\,\rangle_{j}$}
Here we calculate $\langle\,\hat{\chi}_{b}^{2}\,\rangle_{j}$,
\begin{align*}
	 \langle\,\hat{\chi}_{b}^{2}\,\rangle_{j}\,\mathcal{Z}^{(0)}[j]&=\int\!dq_{b}\,dr_{b}\int\!dq_{a}\,dr_{a}\;\delta(q_{b})\left[r_{b}+\frac{1}{2}\,q_{b}\right]^{2}J(q_{b},r_{b},t;q_{a},r_{a},0)\rho_{\chi}(q_{a},r_{a},0)\notag\\
	 &=-\frac{1}{\bigl[m\dot{\beta}(0)\bigr]^{2}}\biggl\{-2I_{22}-\frac{1}{2\sigma^{2}}-\frac{m^{2}\sigma^{2}}{2}\,\dot{\alpha}^{2}(0)\biggr.\notag\\
	 &\qquad\qquad\qquad\qquad\qquad\qquad\qquad\quad+\biggl.\Bigl[-i\,m\,\dot{\alpha}(0)\,b-i\,m\,\dot{J}_{r}(0)-2I_{22}\,\bar{q}_{a}-I_{20}-\frac{1}{2\sigma^{2}}\,\bar{q}_{a}\Bigr]^{2}\biggr\}\notag\\
	 &\qquad\qquad\qquad\times\exp\biggl\{i\,\Bigl(C-\frac{\dot{J}_{r}(0)}{\dot{\beta}(0)}\Bigr)-\frac{\sigma^{2}}{4}\Bigl(A-\frac{\dot{\alpha}(0)}{\dot{\beta}(0)}\,B\Bigr)^{2}-\frac{1}{4\sigma^{2}}\,\bar{q}_{a}^{2}-I_{22}\bar{q}_{a}^{2}-I_{20}\bar{q}_{a}-I_{00}\biggr\}\,,\notag
\end{align*}
with
\begin{equation*}
	\bar{q}_{a}=\frac{B}{m\dot{\beta}(0)}=\frac{1}{m}\int_{0}^{t}\!ds\;D_{2}(s)\,j_{q}(s)\,,\qquad\qquad b=\frac{i}{2}\,\Bigl(A-\frac{\dot{\alpha}(0)}{\dot{\beta}(0)}\,B\Bigr)\,\sigma^{2}=i\,\frac{\sigma^{2}}{2}\int_{0}^{t}\!ds\;D_{1}(s)\,j_{q}(s)\,.
\end{equation*}
From the results in the previous section, we have $\langle\,\chi_{b}^{2}\,\rangle$ given by
\begin{align}
	\langle\,\hat{\chi}_{b}^{2}\,\rangle_{j}&=\frac{1}{2}\left[\sigma^{2}D_{1}^{2}(t)+\frac{1}{m^{2}\sigma^{2}}\,D_{2}^{2}(t)\right]+\frac{e^{2}}{m^{2}}\int_{0}^{t}\!ds\!\int_{0}^{t}\!ds'\;D_{2}(t-s)\,G_{H,\,\beta}^{(\phi)}(s-s')\,D_{2}(t-s')\notag\\
	 &\qquad\qquad+\left(\frac{1}{m}\int_{0}^{t}\!ds\;D_{2}(t-s)\,j_{r}(s)+\frac{i}{2}\int_{0}^{t}\!ds\;\Bigl[\sigma^{2}D_{1}(t)D_{1}(s)+\frac{1}{m^{2}\sigma^{2}}\,D_{2}(t)D_{2}(s)\Bigr]j_{q}(s)\right.\notag\\
	 &\qquad\qquad\qquad\qquad\qquad\qquad\qquad\qquad+\left.i\,\frac{e^{2}}{m}\int_{0}^{t}\!ds\!\int_{0}^{t}\!ds'\;D_{2}(t-s)\,G_{H,\,\beta}^{(\phi)}(s-s')\,\mathfrak{J}_{q}(s')\right)^{2}\,.
\end{align}
In the absence of external sources, this reverts to the well-known result 
\begin{align}\label{E:zzz}
	\langle\,\hat{\chi}_{b}^{2}\,\rangle&=\frac{1}{2}\left[\sigma^{2}D_{1}^{2}(t)+\frac{1}{m^{2}\sigma^{2}}\,D_{2}^{2}(t)\right]+\frac{e^{2}}{m^{2}}\int_{0}^{t}\!ds\!\int_{0}^{t}\!ds'\;D_{2}(t-s)\,G_{H,\,\beta}^{(\phi)}(s-s')\,D_{2}(t-s')\,.
\end{align}

\subsubsection{$\langle\, \hat{p}_{b}\,\rangle_{j}$}
Computation of the expectation value of the conjugate momentum is more complicated because it involves differentiation of the evolutionary operator with respect to the coordinates,
\begin{align*}
	\langle \,\hat{p}_{b}\,\rangle_{j}\,\mathcal{Z}^{(0)}[j]&=-i\int\!dq_{b}\,dr_{b}\int\!dq_{a}\,dr_{a}\;\delta(q_{b})\left[\frac{1}{2}\frac{\partial}{\partial r_{b}}+\frac{\partial}{\partial q_{b}}\right]J(q_{b},r_{b},t;q_{a},r_{a},0)\rho_{\chi}(q_{a},r_{a},0)\\
		 &=-i\,\frac{m}{2\pi\,D_{2}(t_{b})}\int\!dq_{b}\,dr_{b}\int\!dq_{a}\,dr_{a}\;\delta(q_{b})\left[\frac{1}{2}\frac{\partial}{\partial r_{b}}+\frac{\partial}{\partial q_{b}}\right]\exp\biggl\{i\,m\,\dot{\alpha}(t)\,q_{b}\,r_{a}+i\,m\,\dot{\beta}(t)\,q_{b}\,r_{b}\biggr.\\
	 &\qquad\qquad+i\,m\,\dot{J}_{r}(t)\,q_{b}-i\,m\,\dot{\alpha}(0)\,q_{a}r_{a}-i\,m\,\dot{\beta}(0)\,q_{a}r_{b}-i\,m\,\dot{J}_{r}(0)\,q_{a}+i\,A\,r_{a}+i\,B\,r_{b}+i\,C\vphantom{\bigg|}\\
	 &\quad-\biggl.I_{11}\,q_{b}^{2}-I_{22}\,q_{a}^{2}-I_{12}\,q_{b}q_{a}-I_{10}\,q_{b}-I_{20}\,q_{a}-I_{00}\biggr\}\times\left(\frac{1}{\pi\sigma^{2}}\right)^{1/2}\exp\biggl\{-\frac{1}{\sigma^{2}}\left[r_{a}^{2}+\frac{1}{4}\,q_{a}^{2}\right]\biggr\}\notag\\
	 &=-i\biggl\{i\,m\,\dot{\alpha}(t)\,b+i\,m\,\dot{J}_{r}(t)-I_{12}\,\bar{q}_{a}-I_{10}\biggr.\notag\\
	&\qquad\qquad\qquad\qquad\qquad\qquad\quad+\biggl.\frac{\dot{\beta}(t)}{\dot{\beta}(0)}\Bigl[-i\,m\,\dot{\alpha}(0)\,b-i\,m\,\dot{J}_{r}(0)-2I_{22}\,\bar{q}_{a}-I_{20}-\frac{1}{2\sigma^{2}}\,\bar{q}_{a}\Bigr]\biggr\}\\
	 &\qquad\qquad\qquad\times\exp\biggl\{i\,\Bigl(C-\frac{\dot{J}_{r}(0)}{\dot{\beta}(0)}\Bigr)-\frac{\sigma^{2}}{4}\Bigl(A-\frac{\dot{\alpha}(0)}{\dot{\beta}(0)}\,B\Bigr)^{2}-\frac{1}{4\sigma^{2}}\,\bar{q}_{a}^{2}-I_{22}\bar{q}_{a}^{2}-I_{20}\bar{q}_{a}-I_{00}\biggr\}\,.
\end{align*}
We make some re-arrangement
\begin{align}
	\dot{\alpha}(t)-\frac{\dot{\beta}(t)}{\dot{\beta}(0)}\,\dot{\alpha}(0)&=\dot{D}_{1}(t)\,,\\
	 i\,m\Bigl[\dot{\alpha}(t)-\frac{\dot{\beta}(t)}{\dot{\beta}(0)}\,\dot{\alpha}(0)\Bigr]b&=-\frac{m\sigma^{2}}{2}\!\int_{0}^{t}\!ds\;\dot{D}_{1}(t)D_{1}(s)\,j_{q}(s)\,,\\
	 i\,m\Bigl[\dot{J}_{r}(t)-\frac{\dot{\beta}(t)}{\dot{\beta}(0)}\,\dot{J}_{r}(0)\Bigr]&=i\!\int_{0}^{t}\!ds\;\dot{D}_{2}(t-s)\,j_{r}(s)\,,\\
	 -\Big[I_{12}\,\bar{q}_{a}+I_{10}\Bigr]&=-e^{2}\int_{0}^{t}\!ds\!\int_{0}^{t}\!ds'\;\alpha(t-s)\,G_{H,\,\beta}^{(\phi)}(s-s')\,\mathfrak{J}_{q}(s')\,,\\
	 -\frac{\dot{\beta}(t)}{\dot{\beta}(0)}\Big[2I_{22}\,\bar{q}_{a}+I_{20}\Bigr]&=-e^{2}\dot{D}_{2}(t)\int_{0}^{t}\!ds\!\int_{0}^{t}\!ds'\;\beta(t-s)\,G_{H,\,\beta}^{(\phi)}(s-s')\,\mathfrak{J}_{q}(s')\,,\\
	\alpha(t-s)+\dot{D}_{2}(t)\beta(t-s)&=\dot{D}_{2}(t-s)\,.
\end{align}
Thus, the expectation value of the conjugate momentum is
\begin{align}\label{E:fgkjras}
	\langle\, \hat{p}_{b}\,\rangle_{j}&=\int_{0}^{t}\!ds\;\dot{D}_{2}(t-s)\,j_{r}(s)+\frac{i}{2}\!\int_{0}^{t}\!ds\;\left[m\sigma^{2}\dot{D}_{1}(t)D_{1}(s)+\frac{1}{m\sigma^{2}}\,\dot{D}_{2}(t)D_{2}(s)\right]j_{q}(s)\notag\\
	 &\qquad\qquad\qquad\qquad\qquad\qquad\qquad\qquad\quad+i\,e^{2}\int_{0}^{t}\!ds\!\int_{0}^{t}\!ds'\;\dot{D}_{2}(t-s)\,G_{H,\,\beta}^{(\phi)}(s-s')\,\mathfrak{J}_{q}(s')\,.
\end{align}
This reduces to $\langle \,\hat{p}_{b}\,\rangle=0$ in the absence of external sources. Note that $\langle\,\hat{p}_{b}\,\rangle_{j}$ is almost identical to the time derivative of $\langle\,\hat{\chi}_{b}\,\rangle_{j}$, apart from the mass $m$. The latter has an additional term proportional to
\begin{equation}
	\frac{i}{2}\biggl[\sigma^{2}\,D_{1}^{2}(t)+\frac{1}{m^{2}\sigma^{2}}\,D_{2}^{2}(t)\biggr]\,j_{q}(t)\,,
\end{equation}
so we might quickly conclude $\langle\,\hat{p}_{b}\,\rangle_{j}\neq m\langle\,\dot{\hat{\chi}}_{b}\,\rangle_{j}$. This is odd because one expects the equal sign should hold when the oscillator is driven by an external classical source. The resolution lies in the observation that if $j$ describes an external source, then at the end of the calculation, we need to set $j_{q}$ to zero, that is, $j_{+}=j_{-}=j$ exactly like what we do for $\chi_{\pm}$. Doing so we obtain the expected result that $\langle\, \hat{p}_{b}\,\rangle_{j,\,j_{q}=0}=m\langle\,\dot{\hat{\chi}}_{b}\,\rangle_{j,\,j_{q}=0}$.

\subsection{Functional method approach}
{Now, for comparison purpose, we turn to the functional method for a few elementary examples.}
 
\subsubsection{Schwinger function $\langle\,\hat{\chi}(\tau)\hat{\chi}(\tau')\,\rangle$}
Here we use the functional method to compute the Schwinger function
\begin{align}
	\langle\,\hat{\chi}(\tau)\hat{\chi}(\tau')\,\rangle&=\frac{1}{\mathcal{Z}[j;t)}\frac{\delta^{\,2}\mathcal{Z}[j;t)}{\delta j_{-}(\tau)\,\delta j_{+}(\tau')}\;\bigg|_{\substack{j=0\\q=0}}\notag\\
	&=-\frac{1}{\mathcal{Z}[j;t)}\left\{\frac{\delta^{2}\mathcal{Z}[j;t)}{\delta j_{q}(\tau)\,\delta j_{q}(\tau')}-\frac{1}{2}\frac{\delta^{2}\mathcal{Z}[j;t)}{\delta j_{q}(\tau')\delta j_{r}(\tau)}+\frac{1}{2}\frac{\delta^{2}\mathcal{Z}[j;t)}{\delta j_{q}(\tau)\,\delta j_{r}(\tau')}-\frac{1}{4}\frac{\delta^{2}\mathcal{Z}[j;t)}{\delta j_{r}(\tau)\,\delta j_{r}(\tau')}\right\}\;\bigg|_{\substack{j=0\\q=0}}\notag\\
	&=-i\,\frac{\delta\,\Xi[j;\tau')}{\delta j_{q}(\tau)}+\frac{i}{2m}\,D_{2}(\tau'-\tau)-\frac{i}{2m}\,D_{2}(\tau-\tau')+\cdots\;\bigg|_{\substack{j=0\\q=0}}\,,\label{E:djrhber}
\end{align}
where $\cdots$ are terms that will vanish in the limit $j\to0$. Since for the free Klein-Gordon field the Schwinger function can be decomposed into the Hadamard function, the retarded and the advanced Green's functions by,
\begin{equation}
	\langle\,\hat{\phi}(x)\hat{\phi}(x')\,\rangle=G_{H}^{(\phi)}(x,x')-\frac{i}{2}\,G_{R}^{(\phi)}(x,x')+\frac{i}{2}\,G_{A}^{(\phi)}(x,x')\,,
\end{equation}
we arrive at the same identifications of the two-point functions of $\chi$ as in \eqref{E:bgfhke1} and \eqref{E:bgfhke2}, and indeed the functional derivatives \eqref{E:djrhber} may serve as a definition of the Schwinger function for $\chi$.

\subsection{Covariance matrix element $\langle\,\hat{p}^{2}(\tau)\,\rangle$}

When a linear oscillator is coupled with the Klein-Gordon field via coordinate coupling, the conjugate momentum in the reduced system of the oscillator has the form
\begin{equation}
	p(\tau)=m\,\dot{\chi}(\tau)\,.
\end{equation}
We may use this observation and the Schwinger function of $\hat{\chi}$ to find the correlation functions of $\hat{p}(\tau)$ and the covariance matrix elements such as $\langle\,\hat{p}^{2}(\tau)\,\rangle$. For example, the Hadamard function of   $\hat{p}$ can be found directly from the time derivatives of the Hadamard function of $\hat{\chi}$ in \eqref{E:bgfhke1}
\begin{align}
	\frac{1}{2}\langle\,\bigl\{\hat{p}(\tau),\,\hat{p}(\tau')\bigr\}\,\rangle&=\frac{m^{2}}{2}\frac{d^{2}}{d\tau\,d\tau'}\langle\,\bigl\{\hat{\chi}(\tau),\,\hat{\chi}(\tau')\bigr\}\,\rangle\notag\\
	&=\frac{m^{2}}{2}\biggl[\sigma^{2}\dot{D}_{1}(\tau)\dot{D}_{1}(\tau')+\frac{1}{m^{2}\sigma^{2}}\,\dot{D}_{2}(\tau)\dot{D}_{2}(\tau')\biggr]\label{E:bgfhke3}\\
	&\qquad\qquad\qquad\qquad\qquad\qquad+e^{2}\int_{0}^{\tau}\!ds'\!\int_{0}^{\tau'}\!ds''\;\dot{D}_{2}(\tau-s')G_{H,\,\beta}^{(\phi)}(s'-s'')\dot{D}_{2}(\tau'-s'')\,,\notag
\end{align}
where the overdot denotes taking the derivative of a function with respective to its argument. The coincident limit of \eqref{E:bgfhke3} gives
\begin{align}
	\langle\,\hat{p}^{2}(\tau)\,\rangle&=\frac{m^{2}}{2}\biggl[\sigma^{2}\dot{D}_{1}^{2}(\tau)+\frac{1}{m^{2}\sigma^{2}}\,\dot{D}_{2}^{2}(\tau)\biggr]+e^{2}\int_{0}^{\tau}\!ds'\!\int_{0}^{\tau}\!ds''\;\dot{D}_{2}(\tau-s')G_{H,\,\beta}^{(\phi)}(s'-s'')\dot{D}_{2}(\tau-s'')\,.
\end{align}
This is  the well-known result for the momentum dispersion of the free harmonic oscillator coupled to a scalar field.

\subsection{Equal-time commutation relation}

As a consistency check of the functional method, we examine whether the uncertainty relation is satisfied at all times in the sense that 
\begin{equation}\label{E:dbritbcz}
	\langle\bigl[\hat{\chi}(\tau),\,\hat{p}(\tau)\bigr]\rangle=i\,.
\end{equation}
We first find the expectation values of the commutator of $\hat{\chi}$ at two different times. From the Schwinger function of $\hat{\chi}$, we have
\begin{align}
	\langle\bigl[\hat{\chi}(\tau),\,\hat{\chi}(\tau')\bigr]\rangle&=\frac{i}{m}\,\mathfrak{D}_{2}(\tau'-\tau)-\frac{i}{m}\,\mathfrak{D}_{2}(\tau-\tau')\,.
\end{align}
where we have restored the notations $\mathfrak{D}_{i}(s)$ to explicitly show the $\theta(s)$ dependence, namely,  $\mathfrak{D}_{i}(s)=\theta(s)\,D_{i}(s)$.

The lefthand side of \eqref{E:dbritbcz} can then be written as
\begin{align}
	&\quad\langle\bigl[\hat{\chi}(\tau),\,\hat{p}(\tau)\bigr]\rangle=m\lim_{\tau'\to\tau}\frac{d}{d\tau'}\langle\bigl[\hat{\chi}(\tau),\,\hat{\chi}(\tau')\bigr]\rangle\notag\\
	&=i\,\lim_{\tau'\to\tau}\Bigl[\delta(\tau'-\tau)\,D_{2}(\tau'-\tau)+\theta(\tau'-\tau)\,\dot{D}_{2}(\tau'-\tau)+\delta(\tau-\tau')\,D_{2}(\tau-\tau')+\theta(\tau-\tau')\,\dot{D}_{2}(\tau-\tau')\Bigr]\notag\\
	&=i\,.
\end{align}
Thus it shows that the equal-time commutation relation is satisfied at all times when the oscillator interacts with the surrounding scalar field and undergoes nonequilibrium time evolution.

\section{Quantum Statistical Averages in the Nonlinear Model of Sec.~\ref{S:ewbehrfwa}}\label{S:rtbsfgsre}

Here we provide the detailed calculation of the first-order correction to the covariance matrix elements for the nonlinear potential in Sec.~\ref{S:ewbehrfwa}. These results can also be obtained from the functional derivatives of their counterparts in~Appendix~\ref{S:eibde} but may not be as succinct as directly taking the functional derivatives of the in-in generating functional $\mathcal{Z}[j;t)$. We take the latter approach and give two examples of calculating $\langle\,\hat{\chi}_{b}^{2}\,\rangle$ and $\langle\,\hat{p}_{b}\,\rangle$. More terms are presented in arXiv:1912.12803.

\subsection{$\langle\,\hat{\chi}_{b}^{2}\,\rangle$}
For $\langle\,\hat{\chi}_{b}^{2}\,\rangle$, we have the first-order nonlinear contribution $\langle\,\hat{\chi}_{b}^{2}\,\rangle^{(1)}$ given by
\begin{align}
	 \langle\,\hat{\chi}_{b}^{2}\,\rangle^{(1)}&=-i\,\frac{\lambda}{\mathcal{Z}_{V}}\int_{0}^{t}\!d\tau\;\biggl\{\frac{1}{2!}\left(\frac{\delta}{i\,\delta j_{q}(\tau)}\right)^{2}\left(\frac{\delta}{i\,\delta j_{r}(\tau)}\right)+\frac{1}{4!}\left(\frac{\delta}{i\,\delta j_{r}(\tau)}\right)^{3}\biggr\}\notag\\
	 &\qquad\qquad\qquad\qquad\qquad\qquad\qquad\times\biggl\{\langle\,\chi_{b}^{2}\,\rangle_{j=0}^{(0)}+\Xi^{2}[j;t)\biggr\}\,\mathcal{Z}[j;t)\;\bigg|_{j=0}\,.
\end{align}
We break the whole expression into smaller components. E.g.,  for the term containing $(\delta/\delta j_{r})^{3}$, we have
\begin{align}
	&\quad-i\,\frac{\lambda}{\mathcal{Z}_{V}}\int_{0}^{t}\!d\tau\;\biggl\{\frac{1}{4!}\left(\frac{\delta}{i\,\delta j_{r}(\tau)}\right)^{3}\biggr\}\biggl\{\langle\,\chi_{b}^{2}\,\rangle_{j=0}^{(0)}+\Xi^{2}[j;t)\biggr\}\,\mathcal{Z}[j;t)\;\bigg|_{j=0}\notag\\
	&=\frac{\lambda}{4!\mathcal{Z}_{V}}\int_{0}^{t}\!d\tau\biggl\{-\langle\,\chi_{b}^{2}\,\rangle_{j=0}^{(0)}\,\mathfrak{J}^{3}_{q}(\tau)+6i\,\mathfrak{J}_{q}(\tau)\left[\frac{\delta\,\Xi[j;t)}{\delta j_{r}(\tau)}\right]^{2}-6\mathfrak{J}^{2}_{q}(\tau)\,\frac{\delta\,\Xi[j;t)}{\delta j_{r}(\tau)}\,\Xi[j;t)-i\,\mathfrak{J}^{3}_{q}(\tau)\,\Xi^{2}[j;t)\biggr\}\,\mathcal{Z}[j;t)\;\bigg|_{j=0}\notag\\
	&=0\,.\label{E:erhksd}
\end{align}
Similarly we have
\begin{align}
	&\quad-i\,\frac{\lambda}{\mathcal{Z}_{V}}\int_{0}^{t}\!d\tau\;\biggl\{\frac{1}{2!}\left(\frac{\delta}{i\,\delta j_{q}(\tau)}\right)^{2}\left(\frac{\delta}{i\,\delta j_{r}(\tau)}\right)\biggr\}\biggl\{\langle\,\chi_{b}^{2}\,\rangle_{j=0}^{(0)}+\Xi^{2}[j;t)\biggr\}\,\mathcal{Z}[j;t)\;\bigg|_{j=0}\notag\\
	&=\frac{\lambda}{2!\mathcal{Z}_{V}}\int_{0}^{t}\!d\tau\,\biggl\{-\langle\,\chi_{b}^{2}\,\rangle_{j=0}^{(0)}\,\mathfrak{J}_{q}(\tau)\,\frac{\delta\,\Xi[j;\tau)}{\delta j_{q}(\tau)}\,\mathcal{Z}[j;t)-i\,\langle\,\chi_{b}^{2}\,\rangle_{j=0}^{(0)}\,\mathfrak{J}_{q}(\tau)\,\Xi^{2}[j;\tau)\mathcal{Z}[j;t)\biggr.\notag\\
	&\qquad\qquad\qquad\qquad+4i\,\frac{\delta\,\Xi[j;t)}{\delta j_{r}(\tau)}\frac{\delta\,\Xi[j;t)}{\delta j_{q}(\tau)}\,\Xi[j;\tau)\mathcal{Z}[j;t)+2i\,\frac{\delta\,\Xi[j;t)}{\delta j_{r}(\tau)}\frac{\delta\,\Xi[j;\tau)}{\delta j_{q}(\tau)}\,\Xi[j;t)\mathcal{Z}[j;t)\notag\\
	&\qquad\qquad\qquad\qquad-2\,\frac{\delta\,\Xi[j;t)}{\delta j_{r}(\tau)}\,\Xi[j;t)\Xi^{2}[j;\tau)\mathcal{Z}[j;t)+2i\,\mathfrak{J}_{q}(\tau)\left(\frac{\delta\,\Xi[j;\tau)}{\delta j_{q}(\tau)}\right)^{2}\mathcal{Z}[j;t)\notag\\
	&\qquad\qquad\qquad\qquad\qquad\qquad-\biggl.4\,\mathfrak{J}_{q}(\tau)\,\frac{\delta\,\Xi[j;\tau)}{\delta j_{q}(\tau)}\,\Xi[j;t)\Xi[j;\tau)\mathcal{Z}[j;t)-i\,\mathfrak{J}^{2}_{q}(\tau)\,\Xi^{2}[j;t)\Xi^{2}[j;\tau)\mathcal{Z}[j;t)\biggr\}\;\bigg|_{j=0}\notag\\
	&=0\,,\label{E:eorijlqqd}
\end{align}
because all terms that are proportional to $\mathfrak{J}_{q}$ or $\Xi$ vanish in the limit $j_{q}=0=j_{r}$. Therefore, Eqs.~\eqref{E:erhksd} and \eqref{E:eorijlqqd} imply that
\begin{equation}
	\langle\,\hat{\chi}_{b}^{2}\,\rangle^{(1)}=0\,.
\end{equation}

\subsection{$\langle\,\hat{p}_{b}\,\rangle$}
For $\langle\,\hat{p}_{b}\,\rangle$, the leading order nonlinear contribution is given by
\begin{align}
	 \langle\,\hat{p}_{b}\,\rangle^{(1)}&=-i\,\frac{\lambda}{\mathcal{Z}_{V}}\int_{0}^{t}\!d\tau\;\biggl\{\frac{1}{2!}\left(\frac{\delta}{i\,\delta j_{q}(\tau)}\right)^{2}\left(\frac{\delta}{i\,\delta j_{r}(\tau)}\right)+\frac{1}{4!}\left(\frac{\delta}{i\,\delta j_{r}(\tau)}\right)^{3}\biggr\}\;Y[j;t)\mathcal{Z}[j;t)\;\bigg|_{j=0}\,,\notag
\end{align}
where $Y[j;t)$ is given by
\begin{align}
	Y[j;t)&=\frac{i}{2}\!\int_{0}^{t}\!ds\;\left[m\sigma^{2}\dot{D}_{1}(t)D_{1}(s)+\frac{1}{m^{}\sigma^{2}}\,\dot{D}_{2}(t)D_{2}(s)\right]j_{q}(s)+\int_{0}^{t}\!ds\;\dot{D}_{2}(t-s)\,j_{r}(s)\notag\\
	&\qquad\qquad+i\,e^{2}\int_{0}^{t}\!ds\!\int_{0}^{t}\!ds'\;\dot{D}_{2}(t-s)\,G_{H,\,\beta}(s-s')\,\mathfrak{J}_{q}(s')\,.
\end{align}
We now evaluate the leading order nonlinear contribution term by term.

It is straightforward to show
\begin{align}
	-i\,\frac{\lambda}{\mathcal{Z}_{V}}\int_{0}^{t}\!d\tau\;\biggl\{\frac{1}{4!}\left(\frac{\delta}{i\,\delta j_{r}(\tau)}\right)^{3}\biggr\}\;Y[j;t)\mathcal{Z}[j;t)\;\bigg|_{j_{q}=0=j_{r}}&=0\,,
\end{align}
because the functional derivatives of $Y[j;t)$ and $\mathcal{Z}[j;t)$ are respectively given by
\begin{align}
	\frac{\delta\,Y[j;t)}{\delta j_{r}(\tau)}\,&=\dot{D}_{2}(t-\tau)\,,&\frac{\delta^{2}Y[j;t)}{\delta j^{2}_{r}(\tau)}\,&=0\,,\\
	\frac{\delta\,\ln\mathcal{Z}[j;t)}{\delta j_{r}(\tau)}&=\frac{i}{m}\int_{\tau}^{t}\!ds'\;D_{2}(s'-\tau)j_{q}(s')=i\,\mathfrak{J}_{q}(\tau)\,,&\frac{\delta^{2}\ln\mathcal{Z}[j;t)}{\delta j^{2}_{r}(\tau)}&=0\,.
\end{align}
Next we show
\begin{align}
	&\quad-i\,\frac{\lambda}{\mathcal{Z}_{V}}\int_{0}^{t}\!d\tau\;\frac{1}{2!}\left(\frac{\delta}{i\,\delta j_{q}(\tau)}\right)^{2}\left(\frac{\delta}{i\,\delta j_{r}(\tau)}\right)\;Y[j;t)\mathcal{Z}[j;t)\;\bigg|_{j_{q}=0=j_{r}}\notag\\
	&=\frac{\lambda}{2!\mathcal{Z}_{V}}\int_{0}^{t}\!d\tau\,\left(\frac{\delta}{\delta j_{q}(\tau)}\right)^{2}\biggl[\dot{D}_{2}(t-\tau)+i\,\mathfrak{J}_{q}(\tau)\,Y[j;t)\biggr]\mathcal{Z}[j;t)\,\bigg|_{j_{q}=0=j_{r}}\,.\label{E:roeijsjn}
\end{align}
Note that
\begin{equation}
	\frac{\delta}{\delta j_{q}(\tau)}\,\mathfrak{J}_{q}(\tau)=\frac{\delta}{\delta j_{q}(\tau)}\frac{1}{m}\int_{\tau}^{t}\!ds'\;D_{2}(s'-\tau)j_{q}(s')=\frac{1}{2m}\,D_{2}(0)=0\,,
\end{equation}
so the functional derivative $\delta/\delta j_{q}(\tau)$ produces nontrivial results only when it acts on $Y[j;t)$ and $\mathcal{Z}[j;t)$. Since we have
\begin{align}
	\left(\frac{\delta}{\delta j_{q}(\tau)}\right)^{2}\mathcal{Z}[j;t)&=-\langle\,\hat{\chi}^{2}(\tau)\,\rangle_{j=0}^{(0)}\,,&\left(\frac{\delta}{\delta j_{q}(\tau)}\right)^{2}Y[j;t)\mathcal{Z}[j;t)&=0\,,
\end{align}
in the limit $j\to0$, we arrive at
\begin{equation}
	-i\,\frac{\lambda}{\mathcal{Z}_{V}}\int_{0}^{t}\!d\tau\;\frac{1}{2!}\left(\frac{\delta}{i\,\delta j_{q}(\tau)}\right)^{2}\left(\frac{\delta}{i\,\delta j_{r}(\tau)}\right)\;Y[j;t)\mathcal{Z}[j;t)\;\bigg|_{j=0}=-\frac{\lambda}{2!}\int_{0}^{t}\!d\tau\;\dot{D}_{2}(t-\tau)\,\langle\,\hat{\chi}^{2}(\tau)\,\rangle_{j=0}^{(0)}\,,\label{E:rrheieijsjn}
\end{equation}
and thus
\begin{align}
	 \langle\,\hat{p}_{b}\,\rangle^{(1)}&=-\frac{\lambda}{2!}\int_{0}^{t}\!d\tau\;\dot{D}_{2}(t-\tau)\,\langle\,\hat{\chi}^{2}(\tau)\,\rangle_{j=0}^{(0)}=-\frac{m\lambda}{2!}\int_{0}^{t}\!ds\;\dot{G}_{R,0}^{(\chi)}(t-s)\,G_{H,0}^{(\chi)}(s,s)\,,\notag
\end{align}
by the functional method.

\end{document}